\documentclass[10pt,letterpaper]{article}
\usepackage[top=0.85in,left=1.6in,footskip=0.75in]{geometry}

\usepackage{amsmath,amssymb}

\usepackage{changepage}

\usepackage[utf8x]{inputenc}

\usepackage{textcomp,marvosym}

\usepackage{cite}

\usepackage{nameref,hyperref}

\usepackage[right]{lineno}

\usepackage{microtype}
\DisableLigatures[f]{encoding = *, family = * }

\usepackage[table]{xcolor}

\usepackage{array}

\usepackage{soul}

\usepackage{caption,tabularx,booktabs}
\usepackage{pdfpages}

\newcolumntype{+}{!{\vrule width 2pt}}

\newlength\savedwidth

\usepackage[aboveskip=1pt,labelfont=bf,font=small,labelsep=period,singlelinecheck=off]{caption}

\makeatletter
\renewcommand{\@biblabel}[1]{\quad#1.}
\makeatother

\usepackage{lastpage,fancyhdr,graphicx}
\usepackage{epstopdf}
\usepackage{dblfloatfix}
\usepackage{xcolor} 
\usepackage{subcaption}	
\usepackage{multicol}
\usepackage{graphicx}	
\usepackage{mwe}
\usepackage{textcomp}
 \usepackage{steinmetz}

\usepackage[title]{appendix}
\usepackage{algorithm,algpseudocode}
\newcounter{algsubstate}
\renewcommand{\thealgsubstate}{\alph{algsubstate}}
\newenvironment{algsubstates}
  {\setcounter{algsubstate}{0}%
   \renewcommand{\State}{%
     \stepcounter{algsubstate}%
     \Statex {\footnotesize\thealgsubstate:}\space}}

\usepackage{cite}

\usepackage{ifthen}
\usepackage{amsmath}
\usepackage{amssymb}
\usepackage{theorem}
\usepackage{xspace}
\usepackage{calc}
\usepackage{amsfonts}
\usepackage{multicol}
\usepackage{graphicx}	
\usepackage{mwe}
\usepackage{textcomp}
 \usepackage{steinmetz}
 \usepackage{xcolor}

\newcommand{\artworks}{0}
\newcommand{\figures}{1}

\usepackage{ifthen}

\newcommand{\trytoturnpage}{\vspace*{20em}\par\noindent}
\newcommand{\myfig}[1]{\ifthenelse{\artworks=1}{\begin{figure}[f]\trytoturnpage}{\begin{figure}[#1]}}
\newcommand{\mytab}[1]{\ifthenelse{\artworks=1}{\begin{table}[f]\trytoturnpage}{\begin{table}[#1]}}
\newcommand{\myfigstar}[1]{\ifthenelse{\artworks=1}{\begin{figure*}[f]\trytoturnpage}{\begin{figure*}[#1]}}
\newcommand{\mytabstar}[1]{\ifthenelse{\artworks=1}{\begin{table*}[f]\trytoturnpage}{\begin{table*}[#1]}}

\newcommand{\mycaption}[1]{\ifthenelse{\artworks=1}{\vspace*{10em}\caption{#1}}{\caption{#1}}}

\newcommand{\myfigend}{\ifthenelse{\artworks=1}{\trytoturnpage\end{figure}}{\end{figure}}}
\newcommand{\myfigstarend}{\ifthenelse{\artworks=1}{\trytoturnpage\end{figure*}}{\end{figure*}}}

\newcommand{\mytabend}{\ifthenelse{\artworks=1}{\trytoturnpage\end{table}}{\end{table}}}
\newcommand{\mytabstarend}{\ifthenelse{\artworks=1}{\trytoturnpage\end{table*}}{\end{table*}}}

\newcommand{\mycenterwmf}[3]{\ifthenelse{\figures=1}{\centerwmf{#1}{#2}{#3}}{\vskip#2\medskip}}
\newcommand{\myspecial}[1]{\ifthenelse{\figures=1}{\special{#1}}{}}
\newcommand{\mycentereps}[3]{\ifthenelse{\figures=1}{\centereps{#1}{#2}{#3}}{\vskip#2\medskip}}

\newsavebox{\fminibox}
\newlength{\fminilength}
\newenvironment{fminipage}[1][\linewidth]
	{ \setlength{\fminilength}{#1}\addtolength{\fminilength}{-2\fboxsep}%
					       \addtolength{\fminilength}{-2\fboxrule}%
	   \begin{lrbox}{\fminibox}\begin{minipage}{\fminilength}}
	{ \end{minipage}\end{lrbox}\noindent\fbox{\usebox{\fminibox}}}

\newcommand{\gComment}[1]{}

\newcommand{\ttt}{\ensuremath^{\scriptscriptstyle \mathrm{T}}}

\newcommand{\hhh}{\ensuremath^{\scriptscriptstyle \mathrm{H}}}

\newcommand{\vet}[1]{{\rm \bf #1}}

\newcommand{\E}[1]{\mathrm{E}\left\{#1\right\}}

\newcommand{\LLRT}{LLRT\@\xspace}

\newcommand{\ggdef}{\mathop {=} \limits^{\text{def}}}
\newcommand{\geg}{\ensuremath{\!=\!}}

\DeclareMathOperator{\diag}{diag}

{
\theoremheaderfont{\normalfont\scshape}

\newtheorem{Theorem}{Theorem}

}

\newcommand{\ie}{{\textit{i.e.}}\@\xspace}

\newcommand{\sss}[1]{{#1}}

\newcommand{\gn}{n}

\newcommand{\gN}{N}
\newcommand{\gduno}[1][]{\sigma_{1}^{#1}}

\newcommand{\gdN}[1][]{\sigma_{\sssgN}^{#1}}
\newcommand{\sssgN}{\sss{\gN}}

\newcommand{\gdn}[1][]{\sigma_{\gn}^{\vphantom{-}#1}}
\newcommand{\gdnsq}{\gdn[2]}

\newcommand{\gdnm}[1][1]{\sigma_{\gn}^{-#1}}

\newcommand{\gvI}{\vet{I}}
\newcommand{\gD}[1][2]{\boldsymbol{\Sigma}^{#1}}

\newcommand{\geta}{\eta}

\newcommand{\getan}{\eta_{\gn}}
\newcommand{\getansq}{\getan^2}

\newcommand{\gHzero}{\mathcal{H}_0}
\newcommand{\gHuno}{\mathcal{H}_1}

\newcommand{\gvz}{\widetilde{\boldsymbol{l}}}
\newcommand{\gzeta}{\zeta}
\newcommand{\gvzeta}{\boldsymbol{\gzeta}}

\newcommand{\gvxi}{\boldsymbol{\xi}}
\newcommand{\gvx}{\vet{x}}

\newcommand{\gvK}{\vet{K}}
\newcommand{\gvT}{\vet{T}}

\newcommand{\gvKzero}{\vet{K}_0}
\newcommand{\gvKuno}{\vet{K}_1}

\newcommand{\gvQzero}{\vet{Q}_0}

\newcommand{\gvVuno}{\vet{V}_1}
\newcommand{\gveta}{\boldsymbol{\eta}}
\newcommand{\gvLambda}{\boldsymbol{\Lambda}}

\newcommand{\gvLambdauno}{\gvLambda_1}

\newcommand{\gvetazero}{\gveta_0}
\newcommand{\gvetauno}{\gveta_1}

\newcommand{\gvzero}{\vet{0}}
\newcommand{\gNormalLaw}[2]{\mathcal{N}\left({#1},{#2}\right)}

\newcommand{\gx}{x}
\newcommand{\gxuno}{\gx_{1}}
\newcommand{\gxn}{\gx_{\gn}}
\newcommand{\gxN}{\gx_{\sssgN}}
\newcommand{\gup}     [1]{\overset{#1}{\uparrow}}
\newcommand{\gdown}[1]{\underset{#1}{\downarrow}}
\newcommand{\gvrv}[2]{\overset{\gup{#1}}{\underset{\gdown{#2}}{\gtrless}}}

\newcommand{\gP}{P}

\newcommand{\sumdneqone}{\sum_{\gn=1}^{\gP}}
\newcommand{\sumdeqone}{\sum_{\gn=\gP+1}^{\gN}}

\newcommand{\gxi}{\xi}

\newcommand{\gy}{\xi}

\newcommand{\gvy}{\vet{\gy}}

\newcommand{\gmeno}{\!-\!}

\newcommand{\gLRTD}{\mathcal{R}}

\newcommand{\gsigmaysqn}{\sigma_{{\gn}}}

\newcommand{\gsigmaysqnm}{\sigma_{{\gn}}^{-1}}

\newcommand{\getay}{\eta}
\newcommand{\getayn}{\getay_{{\gn}}}

\newcommand{\getaysqn}{\getay^{2}_{{\gn}}}

\newcommand{\gvalpha}{\boldsymbol{\alpha}}

\newcommand{\sumalld}{\sum_{\gn=1}^{\gN}}

\newcommand{\gc}{a}
\newcommand{\gcLLRTn}{\gc_{\mathrm{LLR},\gn}}
\newcommand{\gcLLRTuno}{\gc_{\mathrm{LLR},1}}
\newcommand{\gcLLRTN}{\gc_{\mathrm{LLR},\sssgN}}
\newcommand{\gvcLLRT}{\vet{\gc}_{\mathrm{LLR}}}

\newcommand{\gtildec}{\tilde{\gc}}

\newcommand{\gtildecLLRTnsq}{\gtildec^{2}_{\mathrm{LLR},\gn}}

\newcommand{\gvtildecLLRT}{\vet{\gtildec}_{\mathrm{LLR}}}

\newcommand{\gvtildey}{\tilde{\gvy}}

\newcommand{\gvetatildeyuno}{\boldsymbol{\eta}_{\gvtildey|\gHuno}}
\newcommand{\gvetatildeyzero}{\boldsymbol{\eta}_{\gvtildey|\gHzero}}

\newcommand{\gReal}[1]{\Re\left\{{#1}\right\}}

\newcommand{\gvalphaLLRT}{\gvalpha_{\text{LLR}}}

\newcommand{\galphaLLRTuno}{\alpha_{\text{LLR},1}}

\newcommand{\galphaLLRTNdueN}{\alpha_{\text{LLR},2\gN}}
\newcommand{\galphaLLRTn}{\alpha_{\text{LLR},n}}

\newcommand{\getanconj}{\geta_{\gn}^{\ast}}

\usepackage{lipsum}

\usepackage{textgreek}

\begin{document}
\vspace*{0.2in}

\begin{flushleft}

\newcommand\blfootnote[1]{%
  \begingroup
  \renewcommand\thefootnote{}\footnote{#1}%
  \addtocounter{footnote}{-1}%
  \endgroup
}

{\Large
\textbf\newline{\textbf{Improving J-divergence of brain connectivity states by graph Laplacian denoising}}
}

\bigskip

Tiziana Cattai\textsuperscript{1,2,3},
Gaetano Scarano\textsuperscript{3},
Marie-Constance Corsi\textsuperscript{1,2},
Danielle S. Bassett\textsuperscript{4,5,6,7,8,9}
Fabrizio De Vico Fallani\textsuperscript{1,2},
Stefania Colonnese\textsuperscript{3}
\\
\bigskip
\textbf{1} Inria Paris, Aramis Project Team, Paris, France
\\
\textbf{2} Institut du Cerveau et de la Moelle epiniere, ICM, Inserm U 1127, CNRS UMR 7225, Sorbonne Universite, Paris, France
\\
\textbf{3} Dept. of Information Engineering, Electronics and Telecommunication, Sapienza University of Rome, Italy 
\\
\textbf{4} Department of Bioengineering, University of Pennsylvania, Philadelphia, PA, 19104, USA
\\
\textbf{5} Department of Electrical and Systems Engineering, University of Pennsylvania, Philadelphia, PA, 19104, USA
\\
\textbf{6} Department of Physics \& Astronomy, University of Pennsylvania, Philadelphia, PA, 19104, USA
\\
\textbf{7} Department of Neurology, Hospital of the University of Pennsylvania, Philadelphia, PA, 19104, USA
\\
\textbf{8} Department of Psychiatry, Hospital of the University of Pennsylvania, Philadelphia, PA 19104, USA
\\
\textbf{9} The Santa Fe Institute, Santa Fe, NM, 87501 USA
\\
\bigskip

Corresponding author: tiziana.cattai@uniroma1.it\blfootnote{This work has been submitted to the IEEE for possible publication. Copyright may be transferred without notice, after which this version may no longer be accessible}

\end{flushleft}
\bigskip

\begin{abstract} 
Functional connectivity (FC) can be represented as a network, and is frequently used to better understand the neural underpinnings of complex tasks such as motor imagery (MI) detection in brain-computer interfaces (BCIs). However, errors in the estimation of connectivity can affect the detection  performances. In this work, we address the problem of denoising common connectivity estimates to improve the detectability of different connectivity states. Specifically, we propose a denoising algorithm that acts on the network graph Laplacian, which leverages recent graph signal processing results. Further, we derive a novel formulation of the Jensen divergence for the denoised Laplacian under different states. Numerical simulations on synthetic data show that the denoising method improves the Jensen divergence of connectivity patterns corresponding to different task conditions. Furthermore, we apply the Laplacian denoising technique to brain networks estimated from real EEG data recorded during MI-BCI experiments. Using our novel formulation of the J-divergence, we are able to quantify the distance between the FC networks in the motor imagery and resting states, as well as to understand the contribution of each Laplacian variable to the total J-divergence between two states. Experimental results on real MI-BCI EEG data demonstrate that the Laplacian denoising improves the separation of  motor imagery and resting mental states, and  shortens  the time interval required for connectivity estimation. We conclude that the approach shows promise for the robust detection of connectivity states while being appealing for implementation in real-time BCI applications.
\end{abstract}


\section{Introduction}
\label{sect:Introduction}%

Functional connectivity describes how brain areas mutually interact \cite{bastos2016tutorial}. This information can be modeled as a graph, which is one of the most common formalism to characterize networked data \cite{newman2010networks,torres2020and}. Many recent studies prove that mental states can be characterized by graph statistics, such as node strength, efficiency, and modularity \cite{gonzalez2020network}. Detecting brain connectivity-related features corresponding to different mental states can enhance several technologies, such as brain-computer interfaces (BCIs). BCI systems allow subjects to communicate and interact without peripheral neuro-muscular activity \cite{wolpaw2002brain}. The requirement for the BCI functioning is therefore the correct detection of the user's mental states. While research on the subject has significantly advanced over the last decade, there is still a key limitation known as BCI inefficiency \cite{thompson2019critiquing}. It refers to the fact that there is a percentage of users who cannot be trained to use the interface. This limitation, together with system-user interaction problems \cite{thompson2019critiquing}, motivated us to develop new tools with the intent of having a more robust estimate of brain connectivity with the final goal of better separating two cognitive states. Implementing this estimate from signals acquired at graph nodes (e.g. EEG electrodes) is a difficult task because of the inherent noise, the number of links to estimate, the presence of artifacts, the non-stationarity of the signal. 

To address the problem of connectivity estimation together with the improvement of separability between mental states to optimally control a BCI, it is necessary to combine tools from different fields, such as neuroscience and signal processing. For example, graph signal processing (GSP) can be applied in this scenario \cite{ortega2018graph,sandryhaila2013discrete,shuman2013emerging}. GSP has already been used to deal with biological data, and in particular brain data \cite{huang2018graph,medaglia2018functional}. Indeed, GSP is potentially able to integrate information regarding brain structure, as represented by the graph itself, with information regarding brain function, as represented by the graph signals. 

Another helpful tool to the brain connectivity estimation problem is the signal detection theory. 
Detection procedures can be applied to investigate statistical differences between the brain connectivity features of two different mental states, which corresponds to motor imagery and resting state for our applications. In this context, widely adopted statistics  are the Likelihood Ratio (LR) of the features \cite{scharf1987low,scharf1991statistical,pezeshki2006canonical,trees2001detection,poor2013introduction,kullback1997information,theodoridis2009pattern,schurmann1996pattern} as well as the linear detector  maximizing  the so-called \textit{deflection}   \cite{picinbono1986detection,picinbono1988optimal,picinbono1995deflection,chevalier1996complex}.
With the aim of obtaining a distance metric of features under two states, the case of normally distributed observations simplifies the analysis. Indeed,  for normally distributed observations  with equal conditional variance and different conditional means, the maximum deflection test coincides with the  LR test. Moreover, this latter can be extended to a linear quadratic detector so as to cope with observations characterized by  different conditional variances.  \cite{picinbono1986detection,picinbono1988optimal}. To obtain a measure of separability between features under the two states, one possibility is the Jensen divergence which reflects the maximum deflection test performance \cite{scharf1987low,schurmann1996pattern,pezeshki2006canonical}. 

In the following sub-section, we present our original contributions.

\textbf{Paper Contributions}
\begin{itemize}

\item This paper proposes a novel graph Laplacian denoising algorithm, to enhance the accuracy of brain connectivity estimates. In recent literature, several studies have been conducted to improve the accuracy of link estimation, whereas few studies approach this problem in terms of description of the graph algebraic structure (see Section \ref{sec:relwork}). 
We address this limitation by proposing a subspace-based Laplacian denoising algorithm that preserves relevant connectivity features while rejecting noise-dominated components. In particular, the Laplacian denoising preserves i) the sub-spaces more directly related to the graph topology, as summarized by the eigenvectors corresponding to the smallest Laplacian eigenvalues, and ii) the sub-spaces estimated under a favourable signal-to-noise ratio, as summarized by the eigenvectors corresponding to the largest Laplacian eigenvalues. The noise rejection obtained by this twofold sub-space selection notably improves the separability of two connectivity states. To sake of clarity, we refer to connectivity states as the patterns of connectivity estimated while the brain performs distinct cognitive tasks.
\item In order to measure the improvement achieved by the proposed brain connectivity denoising algorithm, we provide an analysis of the J-divergence of the Laplacian coefficients, we explicit their  contribution to the states' separability in terms of their first and second order moments of the test statistics, and we  show that the proposed Laplacian denoising  actually  increases   the J-divergence of the brain connectivity  features \textit{rest} (\textit{null}) or  \textit{motor imagery}  (\textit{alternative}). The improvement of the J-divergence of the graph   Laplacian coefficients under different connectivity states  is assessed by numerical simulations  on synthetic data. 
\item  Finally, we present experimental results on real EEG data acquired during motor imagery-based BCI experiments, and we prove that the proposed novel denoising algorithm increases  the J-divergence of brain connectivity states and paves the way for connectivity estimation time interval reduction. As a relevant by-product of the theoretical J-divergence analysis, we are able to attribute a score  to each and every Laplacian coefficient representing its marginal  contribution to the J-divergence. The score admits relevant biological interpretation confirming the efficacy of the approach. These results can be assessed by further studies on the brain connectivity features.
\end{itemize}

The structure of the paper is as follows.
Section \ref{sec:relwork} reviews the scientific literature related to our work.  Section \ref{sec:sigmod} describes the signal model used in the analysis. Section \ref{sec:glfilt} details the novel graph denoising we propose. Section \ref{sec:Jdiv} describes the problem of Gaussian detection, and it presents the novel formulation of J-divergence we use throughout the paper. In section \ref{sec:synthres}, we test our filtering method on synthetic graph to verify its ability to separate two graphs. Section \ref{sec:real} applies our method on real EEG data, exhibiting its capacity to estimate graph connectivity during motor imagery tasks and to discriminate between two mental states.
We conclude in section \ref{sec:concl}.
In Table \ref{table:not}, we list of the main notation used in the paper. 
\begin{table}[]
\centering
 \begin{tabular}{||c|c||} 
 \hline
 \textbf{Notation} & \textbf{Description}  \\ [0.5ex] 
 \hline\hline
 $\mathbf{A}$ ,$\mathbf{\hat{A}}$ & adjacency matrix (real, estimated) \\ 
 \hline
 V & set of all nodes  \\
 \hline
 N & total number of nodes  \\
 \hline
 E & set of all links \\
 \hline
 $\mathbf{D}$, $\mathbf{\hat{D}}$ & degree matrix (real, estimated)\\
 \hline
  $\mathbf{L}$, $\mathbf{\hat{L}}$ & Laplacian matrix (real, estimated)\\
 \hline
  $\lambda$, $\hat{\lambda}$ & eigenvalue (real, estimated)\\
 \hline
 $\mathbf{u}$, $\mathbf{\hat{u}}$ & eigenvector (real, estimated)\\
 \hline
  $\mathcal{U_L}$, $\mathcal{U_M}$, $\mathcal{U_H}$ & subset of smallest, central, larger eigenvalues \\ 
 \hline
  $\mathbf{\tilde{L}}$,  $\mathbf{\tilde{l}}$ & filtered graph laplacian matrix and vector\\
 \hline
  $\mathbf{T}$ & transformation matrix\\
 \hline
  $\mathbf{x}$ & vectorized laplacian in the transformed domain\\
 \hline
  J & J-divergence\\
 \hline
  $\mathbf{S}$ & score\\
 \hline
\end{tabular}
\caption{Table of main notation.}
\label{table:not}
\end{table}

\section{Related Work}
\label{sec:relwork}

The problem of graph connectivity estimation has been well studied in literature in different domains, from neuroscience to signal processing and graph theory
\cite{bullmore2009complex,segarra2018statistical,boccaletti2006complex}.
State-of-the-art graph learning methods have the limitation that they usually present over-simplified models for the signal on graph to overcome problems of computational and memory cost. Some recent works, such as \cite{ghoroghchian2020node}, propose different strategies to deal with graph learning problems. Specifically, in the context of mental state identification, authors in \cite{ghoroghchian2020node} present a novel technique to create and modify embeddings associated to each graph node to efficiently compute the adjacency matrix. 
Since FC computation requires a lot of time and computational power, one possibility consists in clustering FC into relevant communities of synchronous components. One approach, recently proposed in \cite{frusque2020multiplex}, goes in this direction with the application of k-means clustering algorithm followed by a tensor decomposition to reduce the FC data.

FC estimation can leverage the generalization of classical signal processing operations into the graph setting, where signals are localized on graph nodes, giving rise to novel research domain of the graph signal processing (GSP) \cite{shuman2013emerging,ortega2018graph,sandryhaila2013discrete}. GSP has already showed its potential to describe brain functioning in \cite{huang2016graph} and \cite{huang2018graph}. Indeed, GSP representation naturally fits to the brain, where the structure can be described by the graph itself while brain functioning directly corresponds to graph signals. An interesting application is  graph filtering \cite{ortega2018graph,rabbat2016graph} which can be useful to extract meaningful brain behaviour \cite{rui2016dimensionality}.  
In \cite{wang2017topographic}, authors propose a mathematical model for brain fibers able to describe neurophysiological mechanisms. The model, based on GSP techniques, extracts a subset of graph eigenvectors which represent a suitable basis for filtering fiber tracts from brain imaging data.

GSP techniques have been applied also to brain-computer interfaces with NIRS signals \cite{petrantonakis2018single}. Specifically, GSP analysis is leveraged in \cite{petrantonakis2018single} in the context of feature extraction to extract spatial information from the NIRS signals and it has been shown to improve classification performances.

Classical signal processing techniques and eigenvector-based filtering have already been used with brain data \cite{spencer1992adaptive,strobach1994event}. In \cite{chen2001eigenvector} and \cite{zhang2005eigenvector}, eigenvector-based filters are applied to fetal magnetic signals and diffuse optic imaging data to obtain more localized activities and reduce artifacts and noise. Specifically, in \cite{zhang2005eigenvector}, classical eigenvector-filtering, i.e. based on larger eigenvectors, is used in diffuse optical imaging with the aim to improving connectivity estimation.

In the following, stemming from the GSP approach to FC estimation, we propose a novel Laplacian denoising algorithm, and we show that it improves the statistical separation of distinguished connectivity states. To this aim,  we provide an analysis of the J-divergence, which naturally provides  a  metric  to quantify the distance between two distributions,  for the problem under concern. Recently,  the J-divergence has been applied  in \cite{li2019time} to investigate the time series' irreversibility . Another recent application of the J-divergence is proposed in \cite{nielsen2019generalization}, where authors present a novel approach to vector-skew the J-divergence. This method is able to preserve J-divergence properties and simultaneously to fine tune parameters for specific applications.

J-divergence has been also applied  in the context of BCI design, to tackle one of the most challenging issues of  EEG-based BCIs, which is the long calibration time.  In general, the number of data required to calibrate the model is really high, because of the noise and the non-stationarity of brain signals.  One solution comes from \cite{giles2019subject}, where a subject-to-subject transfer learning is proposed to improve the classification performance with limited training data. J-divergence is used in a transfer learning framework to test their method by comparing the data of the target subject with the data from previous subjects.
In the following, we investigate the J-divergence under a different points of view, namely i) we assess the performance of the denoising algorithm in separating connectivity states  and ii) we provide criterion for scoring the Laplacian coefficients based on their contribution to the connectivity states separation.


 \section{Signal Model}
 \label{sec:sigmod}

 We are interested in analyzing signals defined on an undirected, connected, weighted graph $G=\{V,E,\boldsymbol{A}\}$, which consists of a finite set of vertices \textit{V} with $\vert V \vert=N$, a set of edges \textit{E} and a weighted adjacency matrix \textbf{A}. If there is an edge $e=(i,j)$ connecting vertices \textit{i} and \textit{j}, the element $A_{i,j}$ represents the weight of the edge; otherwise, $A_{i,j}=0$.
 
 The  graph Laplacian, is a real symmetric matrix  defined%
 \footnote{We refer here to the non-normalized graph Laplacian, also called the combinatorial Laplacian.} as:
\begin{equation}
    \boldsymbol{L}=\boldsymbol{D}-\boldsymbol{A}  
    \label{eq:laplacian}
\end{equation}
 where the degree matrix $\boldsymbol{D}$ is a diagonal matrix whose $i^{th}$ diagonal element $d_i$ is equal to the sum of the weights of all the edges incident to vertex $i$. 
We denote by $\{\boldsymbol{u}_i\}_{i=0,1,...,N-1}$ set of orthonormal eigenvectors, corresponding to increasingly ordered  eigenvalues  $0=\lambda_0\leq\lambda_1\leq\lambda_2...\leq\lambda_{N-1}=\lambda_{max}$. 

In GSP, the Laplacian eigenvectors are considered as SoGs and provide a basis for the Graph Fourier Transform.  For a SoG \textbf{s}, the GFT is defined as the projection of \textbf{s}  on the eigenvectors of the graph Laplacian:
\( \hat{\vet{s}}(\lambda_l)=\vet{s}^H \vet{u}_l
\). The graph Laplacian eigenvalues  $\lambda_l, l=0,\cdots N-1$  have an  analogous meaning to Fourier transform frequency, i.e. smaller  eigenvalues are associated to   eigenvectors that exhibit smoother variations over  connected nodes.
 
In many SoG application problems, including brain functional connectivity estimation, signal values  are actually represented by discrete sequences, obtained   by  sampling  a continuous time signal at each graph node.  

 Let us denote the sequences of samples acquired  over an observation period $T_{oss}$ with sampling pace $T_s$ as $y_n[kT_s],\;n=0,\dots N-1 , k=0,\dots N_s$, $N_s= \lfloor T_{oss}/T_s\rfloor$,  or in vector form as $\boldsymbol{y}[kT_s]=\left[y_0[kT_s]\dots y_{N-1}[kT_s]\right]$
 The vector sequence $\boldsymbol{y}[kT_s],\;k=0,\dots N_s$   is  used  to estimate the graph adjacency matrix $A$ by computing a similarity metric on each and every node pair. There are many state-of-the-art methods to estimate $A_{i,j}, i,j=0,\cdots N\!-\!1$, which associate link weights according to different interaction properties \cite{bastos2016tutorial,friston2011functional,carter1987coherence,nolte2004identifying}. 
 Thereby, the adjacency matrix $A$ is actually represented by its  estimated version $\hat{\mathbf{A}}$, which contains the  connectivity values $\hat{A}_{i,j}$ estimated for each graph node pair $(i,j), i,j=0,\cdots N\!-\!1$. Accordingly, the estimated degree matrix $\hat{\mathbf{D}}$ is computed, so as to  derive the estimated laplacian  $\hat{\mathbf{L}}$ through Eq. \ref{eq:laplacian}, that becomes here:
 \begin{equation}
    \hat{\boldsymbol{L}}=\hat{\boldsymbol{D}}-\hat{\boldsymbol{A}}  
    \label{eq:estlaplacian}
\end{equation}
Any estimation error on the adjacency matrix affects  the Laplacian estimate, and it results into  less distinguishable connectivity states. In the following section we address the denoising of the estimated Laplacian for the purpose of improving the separation of connectivity states.

\section{Graph Connectivity Denoising}
 \label{sec:glfilt}
In order to introduce the Laplacian denoising algorithm,  we consider the  eigenvalue decomposition of the estimated Laplacian matrix $\hat{\mathbf{L}}$ as follows:
\begin{equation}
        \hat{\mathbf{L}} = \sum_{i=0}^{N-1}\hat{\lambda}_i \hat{\mathbf{u}}_i \hat{\mathbf{u}}_i^H
        \label{eq:laplestim}
\end{equation}
Perturbations affect graph Laplacian estimation in terms of both eigenvalues and/or eigenvectors.
To elaborate on the effect of perturbations, we explicit  the first, second and third order error contributions to the  estimated laplacian $\hat{\mathbf{L}}$   as:
\begin{equation}
\begin{split}
        \hat{\mathbf{L}}
        & =\sum_{i=0}^{N-1} (\lambda_i+\epsilon_{\lambda_i})
        (\mathbf{u}_i+\mathbf{\epsilon}_{u_i})
        (\mathbf{u}_i+\boldsymbol{\epsilon}_{u_i})^H
        \\         &=
    \sum_{i=0}^{N-1}    \underbrace{    \lambda_i\mathbf{u}_i\mathbf{u}_i^H    }_{{\boldsymbol{L}}}    +    \underbrace{\lambda_i\mathbf{u}_i\boldsymbol{\epsilon}_{u_i}^H    +\lambda_i    \boldsymbol{\epsilon}_{u_i}      \mathbf{u}_{i}^H +\boldsymbol{\epsilon}_{\lambda_i}\mathbf{u}_i \mathbf{u}_i^H     }_{\textit{first order error}}
    \\&    +
  \underbrace{  \lambda_i \boldsymbol{\epsilon}_{u_i}\boldsymbol{\epsilon}_{u_i}^H  +\epsilon_{\lambda_i}\mathbf{u}_i\boldsymbol{\epsilon}_{u_i}^H+  \epsilon_{\lambda_i}\boldsymbol{\epsilon}_{u_i}\mathbf{u}_i^H  }_{\textit{second order error}}
  +\underbrace{  \epsilon_{\lambda_i}\boldsymbol{\epsilon}_{u_i}\boldsymbol{\epsilon}_{u_i}^H  }_{\textit{third order error}}
\end{split}
\label{eq:laplerror}
\end{equation}

Thereby, the estimated Laplacian can be  approximated at the first order  as the sum of $N$ terms:

\begin{equation}
    \begin{split}
         \hat{\mathbf{L}} \approx \sum_{i=0}^{N-1}(\lambda_i +\epsilon_{\lambda_i})\boldsymbol{u}_i\boldsymbol{u}_i^H + \lambda_i(\boldsymbol{u}_i\boldsymbol{\epsilon}_{u_i}^H+\boldsymbol{\epsilon}_{u_i}\boldsymbol{u}_i^H)
    \end{split}
\label{eq:laplfirst}
\end{equation}

 Eq.\eqref{eq:laplfirst} highlights  the first order error contribution due to relative perturbation of the Laplacian  eigenvalues as well as of the  eigenvectors direction.  We are interested in  the Laplacian components whose  perturbation is contained because either the relative eigenvalue perturbation $\epsilon_{\lambda_i}/\lambda_i$  or the eigenvector perturbation $\epsilon_{u_i}$ is (relatively) small. To this aim, we  order the set of orthonormal eigenvectors $\mathcal{U}_{ALL}=\big\{\hat{\boldsymbol{u}}_l, {l=0,1,...,N-1}\big\}$  with increasingly eigenvalues $0=\hat{\lambda}_0\leq\hat{\lambda}_1\leq\hat{\lambda}_2...\leq\hat{\lambda}_{N-1}:=\hat{\lambda}_{max}$, and we consider three subsets of eigenvalues and associated eigenvectors: $1)$ the subset $\mathcal{U}_L$ containing the  $N_L$ smallest eigenvalues;  $2)$ the subset $\mathcal{U}_H$ containing the $N_H$ largest eigenvalues; and $3)$ the subset $\mathcal{U}_M$ containing the remaining $N_M=N-N_L-N_H$ central eigenvalues, with $\mathcal{U}_L\cup \mathcal{U}_M \cup \mathcal{U}_H=\mathcal{U}_{ALL}$. 

 Firstly, we remark that the $N_H$ largest eigenvalues  are more robust to eigenvalue perturbation; this assumption is usually exploited in classical signal processing, where the subspace  $\mathcal{U}_H$ is used  for the estimation of the covariance matrix because of its favorable  signal-to-noise ratio \cite{ollila2012complex}. 
 
 Secondly, stemming on recent literature results   \cite{ceci2020graph}, it can be expected that the subspace  $\mathcal{U}_L$ is  partially robust in terms of eigenvector perturbation.
  In fact, in \cite{ceci2020graph} the authors states that a connectivity estimation error on the  $A_{mn}$ adjacency matrix element, i.e. on the weight of the link between the $m$-th and the $n$-th nodes, induces a    perturbation $\boldsymbol{\epsilon}_{u_i}$ of the $i$-th eigenvector  depending on the difference between the $m$-th and the $n$-th coefficients of $\mathbf{u_i}$.  Thereby, eigenvectors smoothly varying across the $m$-th and the $n$-th nodes  are less affected by  estimation errors on $A_{mn}$. On the other hand, in GSP, it is well known that  $\mathcal{U}_L$ eigenvectors, corresponding to small eigenvalues, represent low frequency basis elements in the Graph Fourier Transform \cite{von2007tutorial},\cite{shuman2013emerging} since they are  characterized by the smallest variations across strongly connected graph regions.  Thereby, the eigenvectors in $\mathcal{U}_L$  are equipped with inherent resilience to connectivity estimation error within these regions. 
  On the other hand, the  eigenvectors in $\mathcal{U}_L$ are tightly related to the network connectivity, and therefore they need to be involved in the devised denoising method.  
  
  Stemming on these observations, we propose a  denoising method based on preserving the contribution to the Laplacian due to the  subspaces $\mathcal{U}_L$, $\mathcal{U}_H$  while  discarding  those relative to the subspace  $\mathcal{U}_M$. In formulas, given the estimated Laplacian
  \begin{equation}
       \hat{\mathbf{L}} = \sum_{i\in \mathcal{U}_L \cup \mathcal{U}_M \cup \mathcal{U}_H}\hat{\lambda}_i \hat{\mathbf{u}}_i \hat{\mathbf{u}}_i^H 
       \label{eq:lapleig}
\end{equation}
we compute the denoised Laplacian $\tilde{\mathbf{L}}$ as follows:
\begin{equation}
        \tilde{\mathbf{L}} = \sum_{i\in \mathcal{U}_L}\hat{\lambda}_i \hat{\mathbf{u}}_i \hat{\mathbf{u}}_i^H + \sum_{i\in \mathcal{U}_H}\hat{\lambda}_i \hat{\mathbf{u}}_i \hat{\mathbf{u}}_i^H
        \label{eq:lapleigdenoise}
\end{equation}

 We recognize that the proposed, subspace-based, Laplacian denoising approach allows preservation of  
\begin{itemize}
    \item the sub-space $\mathcal{U}_L$,  which is  directly related  to the graph connectivity features
    \item the sub-space $\mathcal{U}_H$, which is estimated with a favourable signal-to-noise ratio.
\end{itemize}

The proposed Laplacian denoising method, synthetically presented in Algorithm $1$, preserves the information   relevant for the purpose of graph connectivity identification, while  rejecting noisy components. In order to quantify the improvement achieved in terms of connectivity state separability, we resort to  the J-divergence   as a metric of the  distance between  connectivity states. In the following  we derive a  formulation the J-divergence  for the problem under concern.

\section{Jensen divergence of connectivity states}
\label{sec:Jdiv}%
 Several metrics can be adopted to determine the separability of two connectivity states \cite{basseville2013divergence}, as represented by the Laplacian matrix $L$.
 Herein, we resort to the notion of J-divergence for characterizing the separability of  connectivity states, and  we reformulate it for the problem under concern. Thus, J-divergence is used to   identify the  Laplacian coefficients  that are most relevant for detection purposes and later on to measure the improvement achieved by the proposed  denoising algorithm.

For the purpose of the analysis, we will assume that the Laplacian coefficients  obtained at the output of the denoising algorithm are normally distributed. 
Let us remark that the Gaussian assumption stands in many applications%
\footnote{The reason why this occurs is that the Gaussian assumption tightly models laplacian diagonal elements, computed in each row as the sum of extradiagonal elements in that column, as well as  extradiagonal elements which are often computed as the result of  correlation estimates.},
including  the case of connectivity estimates carried out on real brain signals, and thereby it is often assumed in the literature, e.g. for  Laplacian estimation purposes \cite{ortega2018graph}
. 
Specifically, we assume  that the vector $\tilde{\mathbf{l}}=\text{Vec}(\tilde{L})$ 
is distributed according  to a  multidimensional Gaussian  probability  whose mean vector and covariance matrix differ  under two   different connectivity states,  referred to as the \textit{null}   and  the \textit{alternative} hypotheses \(\gHzero\), \(\gHuno\) in the following: %
\footnote{The notation \(\gvz \sim \gNormalLaw{\gveta_{\mathrm{{j}}}}{\gvK_{\mathrm{j}}}\), with $j\in \{0,1\}$ indicates that the random vector \(\gvz\) is
 Gaussian distributed with mean vector \(\gveta_{\mathrm{{j}}}\) and covariance matrix \(\gvK_{\mathrm{{j}}}\).}
\begin{equation}
\left\{
\begin{array}{ll}
     \gHzero: \gvz &\sim \gNormalLaw{\gvetazero}{\gvKzero} \\
     \gHuno: \gvz &\sim \gNormalLaw{\gvetauno}{\gvKuno}
\end{array}
\right.
\label{eq:Hypotheses}
\end{equation}
As an information theoretic measure of distance between $\tilde{\mathbf{l}}$ under \(\gHzero\) and \(\gHuno\), we now compute in analytical form     
  the {J-divergence},
  which is defined as the expected value of the difference of the Log Likelihood Ratio under the two hypothesis \(\gHzero\) and \(\gHuno\)  \cite{pezeshki2006canonical}. 
   The {J-divergence} formulation will allow us to evaluate to which extent  the connectivity states represented by the Laplacian coefficients are distinguishable from each other. 

\begin{figure}[ht]
		\centerline{\includegraphics[scale=.45]{{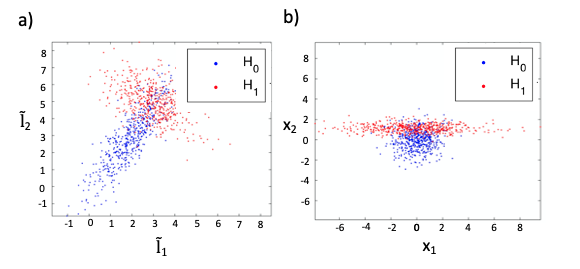}}}
		\caption{Example of transformation effect. In a) we have $2$-dimensional Gaussian distribution which differ under mean and covariance matrix . In b) we report the same distributions after the T transformation}
		\label{fig:jtrasf}
\end{figure}

Let us first assume  that the Laplacian moments $ {\gvetazero}$,${\gvetauno}$,${\gvKzero}$,${\gvKuno}$ are known. 
Detection can then be conducted on a linear transformation of the observations:
\[\gvx \geg \gvT\left(\gvz\gmeno\gvetazero\right)\]
where \(\gveta\ggdef 
\gvT\left(\gvetauno\!-\!\gvetazero\right) \) and \(\gvT=\mathcal{T}({\gvKzero}, \gvKuno)\) is an affine transform  that simultaneously%
\footnote{The matrix \(\gvT\) and the diagonal matrix \(\gD[2]\ggdef\diag(\gduno[2], \dotsc, \gdn[2] , \dotsc, \gdN[2])\) 
are computed as the  \textit{generalized eigenvectors} and  the \textit{generalized eigenvalues} matrices
 of the pencil \((\gvKuno,\gvKzero)\), respectively. Given any square root \(\gvQzero\) of \(\gvKzero^{-1}\), \ie such that \(\gvQzero\hhh\cdot\gvKzero\cdot\gvQzero\geg\gvI\), 
we may conveniently employ the  unitary transformation \(\gvVuno\)
obtained from the eigenanalysis  \(\gvQzero\hhh\cdot\gvKuno\cdot\gvQzero =\gvVuno \cdot \gvLambdauno \cdot \gvVuno\hhh\);
in fact, it is easily proved that the matrix  \(\gvT=\gvVuno\hhh\cdot\gvQzero\hhh\)
verifies \(\gvT \cdot \gvKzero \cdot \gvT\hhh \geg \gvI
\quad;\quad
\gvT \cdot \gvKuno \cdot \gvT\hhh \geg \gD\\
\) with \(\gvLambdauno\geg\gD\).}
 whitens the observations in the \(\gHzero\) hypothesis obtains uncorrelated observations in the \(\gHuno\) hypothesis. 
 An example of the action of the transform $T$, is shown in Fig. \ref{fig:jtrasf} for the case of $2$-dimensional  Gaussian data  whose   mean and covariance matrix differ under  the \(\gHzero,\gHuno\). The original data points are plotted in Fig. \ref{fig:jtrasf}(a), whereas their transformed counterparts are represented in Fig. \ref{fig:jtrasf}(b). The transformed data are unitary variance, zero-centered under \(\gHzero\) and are  uncorrelated  under \(\gHuno\).

 In real detection systems, the moments $ {\gvetazero},{\gvetauno},{\gvKzero},{\gvKuno}$ can be either estimated from a training set, e.g. during a BCI training phase, or coarsely initialized and tracked throughout the system life, using  methodologies, priors, and heuristics related to the application-specific problem under concern \cite{chang2010semiparametric, guttorp2013covariance, greco2014maximum}. Besides, the transformed data $\gvx $  can  be obtained even avoiding  computation of the moments and of $T$, by applying  the laplacian coefficients $\tilde{\mathbf{l}}$ to a suitably trained network \cite{ali2020biometricnet}, that  will enforce the afore-mentioned statistical constraints.  
 
With these position,
the observation model becomes:
\begin{equation}
 \gHzero: \gvx \sim \gNormalLaw{\gvzero}{\gvI}
~\text{versus}~
\gHuno: \gvx \sim \gNormalLaw{\gveta}{\gD}
\label{eq:HypothesesTer}
\end{equation}
The J-divergence is then defined as: 
\begin{equation}
J   \ggdef \E{\gLRTD(\gvx)|\gHuno} - \E{\gLRTD(\gvx)|\gHzero}
\label{eq:jdiv}
\end{equation} being $\gLRTD(\gvx)$ the Log-Likelihood Ratio%
\footnote{The Log-Likelihood Ratio 
\(\gLRTD(\gvx)\) is widely
 adopted  classical LLR detection: $\gLRTD(\gvx)\gvrv{\gHuno}{\gHzero} \theta$, being $\theta$ selected according to the desired detection versus missing probability tradeoff.}:
\begin{equation}
\begin{split}
\gLRTD(\gvx)&=  \gvx\hhh \left(\gvI-\gD[-2]\right)\gvx + 2\gveta\hhh\gD[-2]\gvx
\label{eq:LRTsimple}
\end{split}
\end{equation}


Let us now 
associate the variables \(\gxn\) whose variance \(\gdnsq\neq 1\)
to the first $\gP$ indexes  and the remaining ones to the indexes \(\gn\geg\gP + 1,\dotsc,\gN\)%
\footnote{Possibly, we might have \(P\geg N\) or \(P\geg0\).} so as to rewrite the LLR as follows: 
\newcommand{\LD}{\mathcal{D}}%

\newcommand{\DP}{D}%
\newcommand{\DL}{D_{\mathrm{L}}}%

\begin{equation}
\begin{split}
\gLRTD(\gvx) 
& =\sumalld \frac{1}{\gdnsq} \left[ \left(\gdnsq-1\right) \gxn^2 + 2 {\getan\cdot \gxn}\right]
\\
&  = \sumdneqone \dfrac{\left(\gdnsq-1\right) |\gxn|^2  +  2 {\getan\cdot\gxn}}{\gdnsq}
\\ & \hphantom{=}
+ \!\! \sumdeqone 2{\getanconj \cdot\gxn}
\end{split}
\label{eq:OrderedLRTLQD}
\end{equation}
By adding and subtracting the term 
\({|\getansq|}/{[\gdnsq(\gdnsq-1)]} \) we rearrange the summation  \eqref{eq:OrderedLRTLQD} as:
\begin{equation}
\gLRTD(\gvx) = \underbrace{\sumdneqone\dfrac{\gdnsq-1 }{\gdnsq}  \left|\gxn + \frac{\getan}{\gdnsq-1} \right|^2}_{\textit{\(\gP\) quadratic terms}} 
+ 
 \underbrace{\sumdeqone\!\!2{\getan\cdot \gxn}}_{\textit{\(\gN\gmeno\gP\) linear terms}}
\underbrace{- \sumdneqone \frac{|\getansq|}{\gdnsq \left(\gdnsq-1\right)}   }_{\textit{constant to be included in the threshold}}
\label{eq:SquareCompletationBis}
\end{equation}
The  \(\gP\)  variates \(x_n,\;n\geg1,\dotsc,\gP\),
having different conditional  variances     under the hypotheses \(\gHzero,\gHuno\),
contribute to the LLR by the \(\gP\)  terms quadratic terms. The  \(\gN\gmeno \gP\) variates \(x_n,~n\geg \gP+1,\dotsc,\gN\), 
having  equal unitary  conditional variances under the hypotheses \(\gHzero,\gHuno\), 
contribute to  the LLR by the  \(\gN\gmeno\gP\) linear terms.
To gain further insight on the J-divergence, we resort to the following theorem, whose demonstration is reported in the Appendix.
{\begin{Theorem}
Let $ {\gvxi} $ be a vector formed by the \(N\) statistically independent random variables:
\begin{equation}
\begin{split}
\gxi_n&=  \left(\gxn + \frac{\getan}{\gdnsq-1} \right)^2,\;
\gn=1,\dotsc, \gP\\
\gxi_n&=\gxn, \;\gn=\gP+1,\dotsc, \gN
\end{split}
\label{eq:x2xi}
\end{equation}
The LLR is  expressed as \(
\gLRTD(\gvx)=
{\gvcLLRT\hhh\cdot {\gvxi} }
\)
being  $\gvcLLRT$ constant coefficients defined as in Eq.\eqref{eq:LLRxi}
\label{th:xianda}
and the J-divergence in Eq.\eqref{eq:jdiv} is computed as follows: 
\begin{equation}
\begin{split}
J  
&=\sumdneqone\left(\gdn\gmeno\gdnm\right)^2
\left[1+\dfrac{|\getaysqn|}{\gdn}\dfrac{\gdn+\gdnm}{\left(\gdn-\gdnm\right)^2} \right]
\!+ \!\!\!\sumdeqone \!\!2 |\getansq|
\\&=\sum_{n=1}^{P}J_n^{(\sigma,\eta)}+\sum_{n=P+1}^{N}J_n^{(\eta)}
\label{eq:JDivCosBis}
\end{split}
\end{equation}
\end{Theorem}}
Theorem \ref{th:xianda}  generalizes the result in \cite{scharf1991statistical,pezeshki2006canonical} where only the case of variables having equal conditional means and different  covariances (i.e. ${{\gvetauno}=\gvetazero},\;{{\gvKuno}\neq \gvKzero}$) has been addressed.

The J-divergence as formulated in Eq.\eqref{eq:JDivCosBis} is a measure of the statistical distance of the  graph Laplacian coefficients under two connectivity states   \(\gHuno\) and \(\gHzero\), and it will be used to quantify the improvement of separability of brain connectivity states achieved  by the denoising algorithm described in section \ref{sec:glfilt}.

\begin{figure}[ht]
		\centerline{\includegraphics[scale=.45]{{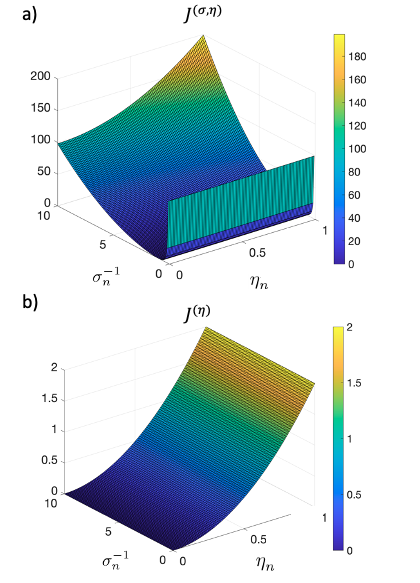}}}
		\caption{J-Divergence contributions as function of mean $\eta$ and standard deviation $\sigma$:  a)  $J^{(\sigma,\eta)}$ for variables whose conditional standard deviation differ under $\gHuno$ and $\gHzero$ , and b)  $J^{(\eta)}$  for variables with invariant conditional standard deviation.}
		\label{fig:j3d}
\end{figure}

Furthermore, the above analysis sheds a light upon the  variables that mostly contribute to the states separability. From Eq.\eqref{eq:JDivCosBis}, we see a one-to-one correspondence between the transformed space variables   $\boldsymbol{x}_n$  and the terms of the J-divergence $J$; besides, the  term  can be of two kinds 
\begin{equation}
    \begin{split}
&J_n^{(\sigma,\eta)}= \left(\gdn\gmeno\gdnm\right)^2
\left[1+\dfrac{|\getaysqn|}{\gdn}\dfrac{\gdn+\gdnm}{\left(\gdn-\gdnm\right)^2} \right]\\&J_n^{(\eta)}=2 |\getansq|
\end{split}\end{equation}
depending on whether  the variable  changes  both in conditional mean and standard deviation, or in conditional mean only.
The functions $J_n^{(\sigma,\eta)}, J_n^{(\eta)}$  are plotted in Fig.\ref{fig:j3d} for $\eta$ between $0$ and $1$ and  $\sigma^{-1}$ from $0$ to $10$.
Interestingly, in Fig.\ref{fig:j3d} we recognize that a conditional variance change   gives a higher contribution to $J$ than an equal  conditional mean change. 

To sum up, the above analysis highlights  the contribution   of each of transformed  variable to the separability of the connectivity states, and allows to rank their relevance to $J_n$, so as to identify the transformed variables which mostly differ under the two hypothesis.  This paves the way for an information theoretic scoring of the Laplacian coefficients, described in the following.

\subsection{J-Divergence based Laplacian coefficients scoring}

As a by-product of the analysis, we are now able to identify which  Laplacian coefficients (i.e. links weights or nodes degrees), contribute mostly  to the connectivity states separability.  This is obtained by attributing a score to the Laplacian coefficients measuring their contribution to the  J-divergence $J$.

By definition, the Laplacian coefficient $\tilde{l}_{\overline{n}},\;\overline{n}=0, \cdots N-1$,  contribute to each  transformed variable $x_n$. 
 Thereby,     we introduce a score $\boldsymbol{S}_{\overline{n}},$,  evaluated by suitable backpropagation of the $J_n$ terms on each contributing Laplacian coefficient. Specifically, the $\overline{n}$-th coefficient score is computed as
\begin{equation}
S_{\overline{n}}=\sum_{n}J_n\cdot\frac{t_{n\overline{n}}}{\sum_{k} t_{nk}} 
\label{eq:score}
\end{equation}
where we recognize that the weight $t_{n\overline{n}}$ representing the contribution of the $\overline{n}$-th Laplacian  coefficient to the ${n}$-th transformed variable is normalized with respect to the sum $\sum_{k} t_{nk}$ of the weights of  all the contributing coefficients.

\begin{figure}[ht]
		\centerline{\includegraphics[scale=.37]{{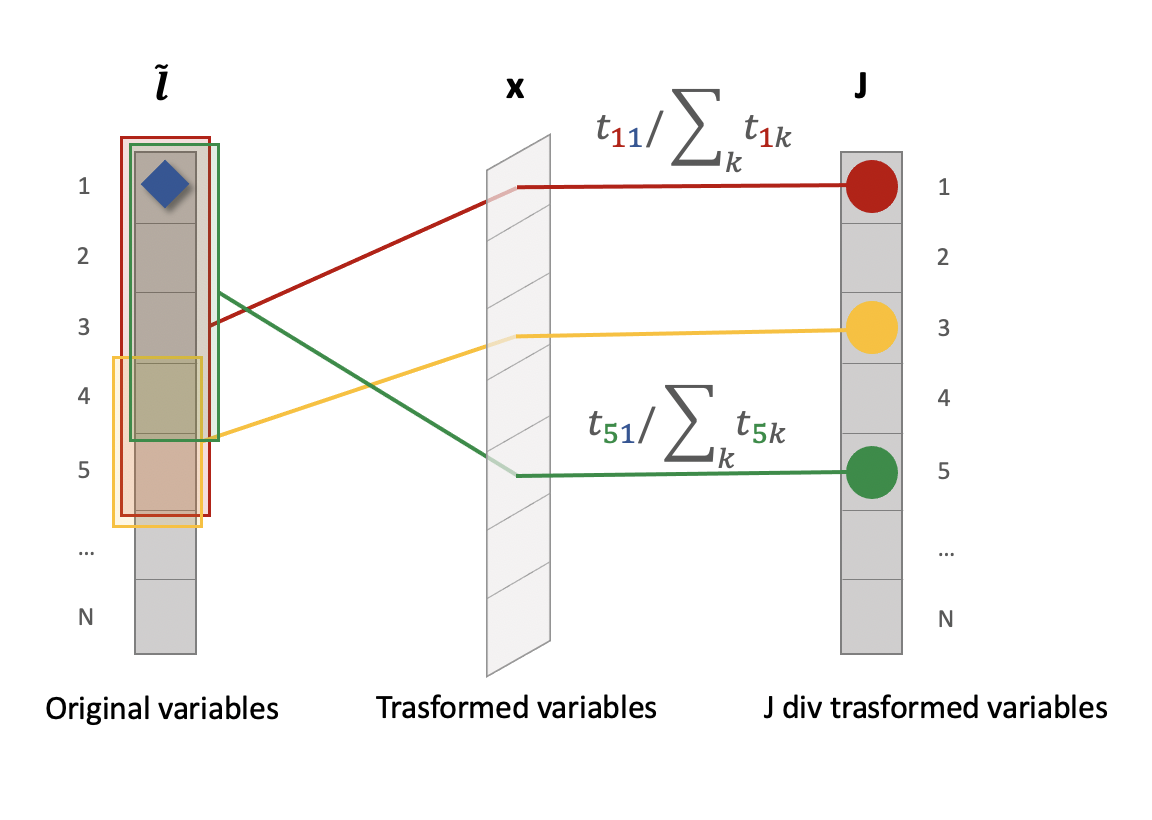}}}
		\caption{Graphic interpretation of the score  computed for the first element in the vector $\tilde{l}$}
		\label{fig:backprop}
\end{figure}

A graphical interpretation of the score  is  provided in Fig.\ref{fig:backprop}, where we represent the set of variables $\tilde{\vet{l}}$ belonging  to the original  domain (left),   the set of  variables $\vet{x}$  belonging  to the transformed  domain (center) and the corresponding  marginal contributions 
to $J$ (right). The relationship between  $\tilde{\boldsymbol{l}}$ and $\boldsymbol{x}$ is given by the transformation matrix $\boldsymbol{T}$ that blends variables from the original domain to  the transformed one. 
Each J-divergence component $J_n$ (colored circle on the right)  is  associated to the variable $\boldsymbol{x_n}$  in the transformed domain, which in turn is originated by many $\tilde{\boldsymbol{l}}_n$ (shaded colored box on the left).   Thereby,  $J_n$    is backprojected to the original domain  by  weighting its contribution as in Eq. \ref{eq:score}. Back-projection and accumulation can also be applied by limiting the summation in Eq.\eqref{eq:score} to the largest  ranking $J_n$  terms. The score computation allows to quantify the relevancy of the Laplacian coefficients $\tilde{\vet{l}}_{\overline{n}}$ for state separability, and  in the experimental results we show  that it leads to meaningful results in case of real BCI data.   
The Algorithm $2$ review the main steps of the J-divergence computation and  scoring  procedures.

\begin{algorithm}[t]
\caption{\textbf{Graph Laplacian denoising}}
\textbf{Input:}  Estimated Laplacian ${\tilde{L}}$
 \\
\textbf{Output:}  Denoised Laplacian ${\tilde{L}}$  \\
\begin{algorithmic}[1]
\State 
Compute the eigen-decomposition $$\hat{L} = \sum_{i=0}^{N-1}\hat{\lambda}_i \hat{\mathbf{u}}_i \hat{\mathbf{u}}_i^H$$
\State 
Compute the denoised Laplacian $\tilde{L}$ by
\begin{algsubstates}

        \State
        Selecting the number $N_{L}$, of smallest eigenvalues and the number $N_H$ of largest eigenvalues to retain
        \State
        Computing  $$\tilde{L}  = \sum_{i=0}^{N_L-1}\hat{\lambda}_i \hat{\mathbf{u}}_i \hat{\mathbf{u}}_i^H +\sum_{i=N-N_H}^{N-1}\hat{\lambda}_i \hat{\mathbf{u}}_i \hat{\mathbf{u}}_i^H 
$$
\end{algsubstates}

\end{algorithmic}
\end{algorithm}

\begin{algorithm}[t]
\caption{\textbf{J-divergence and score computation}}
\textbf{Input:}   Conditional means $\mu_0$, $\mu_1$ and covariance matrices $K_0$, $K_1$ of $\mathbf{\tilde{l}}=\text{Vec}(\tilde{l})$ under $\gHuno$ and $\gHzero$
 \\
\textbf{Output:}  $J_n, \;{S_n},\;n=0,\cdots N-1$ \\
\begin{algorithmic}[1]
\State Step 1: Transform computation
\begin{algsubstates}
\State
Compute the square root matrix $$Q_0\gets K_0^{-1/2}$$
and the eigenvectors  $V_1$ and the eigenvalues $\sigma^2_0,\cdots \sigma^2_{N-1}$ of the eigen-decomposition $$Q_0^H K_1Q_0=V_1 \diag \left(\sigma^2_0,\cdots \sigma^2_{N-1}\right)V_1^H$$ 
\State
Compute  T as $\gets V_1^HQ_0^H$ and $\Sigma\gets \sqrt{(eig(Q_0^HK_1Q_0)')}$
\end{algsubstates}
\State Step 2: J-divergence computation 
\begin{algsubstates}
\State Define a threshold $\theta$  
\State Compute $J_n,\;n=0,\cdots N-1$ as
$$J_n\gets\left\{
\begin{array}{l}
    2 |\getansq|,\; \iff \left(\sigma^2_n>\theta \right)\cup \left(|\sigma^2_n-1|>\theta\right)\\
   \left(\gdn\gmeno\gdnm\right)^2
\left[1+\dfrac{|\getaysqn|}{\gdn}\dfrac{\gdn+\gdnm}{\left(\gdn-\gdnm\right)^2} \right],\:\text{otherwise}
\end{array}
\right.$$, 
\end{algsubstates}

\State Step 3: Score computation
\begin{algsubstates}
\State
Compute $S_{\overline{n}},\;\overline{n}=0,\cdots N-1$ as
$$S_{\overline{n}}=\sum_{n}J_n\cdot\frac{t_{n{\overline{n}}}}{\sum_{k} t_{nk}} $$
\end{algsubstates}

\end{algorithmic}
\end{algorithm}

\section{Results on synthetic data}
 \label{sec:synthres}
 
In this section, we test the performance of the  Laplacian denoising presented in section \ref{sec:glfilt} in improving the J-divergence of two estimated  connectivity states over synthetic SoGs. 
To this end, we first consider a graph and a model for signals at nodes under two connectivity states, selected to represent an over-simplified model of brain EEG signals functional connectivity;  real brain signals are considered in the next section.

We compare our approach with the case of laplacian without filtering (that we refer as $\mathcal{U}_{ALL}$ and with several of eigenvector-based filters, i.e. $\mathcal{U}_{L}$, $\mathcal{U}_{H}$).  Then, we explain in detail the analysis we performed and the related results.

\subsection{ Signal on Graph  generation and connectivity estimation}
In order to validate our framework on synthetic data, we define signals under the two hypothesis $\gHzero$ and $\gHuno$ to obtain two distinct graph connectivity states.  

Under  $\gHuno$, we model the network activity  by considering $H$ scalar generator signals $s^{(h)}[kT_s], h=0,\cdots H-1$. Each generator signal simultaneously contributes to the signals measured over a subset $\mathcal{G}^{(h)}, h=0,\cdots H-1$ of nodes identified by the non-zero component of the $N \times 1$ binary vector $\vet{g}^{(h)}, h=0,\cdots H-1$. 
A noise component $\mathbf{w}[kT_s]$  and a common  component  across all the nodes $b[kT_s]\cdot\mathbf{1}$ are also present. Under  $\gHzero$, only these latter components are observed. 
With these positions, we come up with the following expression for the vector of the observed signals ${\mathbf{y}[kT_s]}$ under the two hypothesis   $\gHuno$ and $\gHzero$:
\begin{equation}
\begin{split}
\gHuno: {\mathbf{y}[kT_s]}&=\sum_{h=0}^{H-1} s^{(h)}[kT_s]\cdot{\mathbf{g^{(h)}}}  +{\mathbf{w}[kT_s]}+b[kT_s]\cdot{\mathbf{1}}
\\
\gHzero: \mathbf{y}[kT_s]&=\mathbf{w}[kT_s]+b[kT_s]\cdot{\mathbf{1}}
\label{eq:modelH0H1}
\end{split}
\end{equation}

In the simulations, the noise $\mathbf{w}[kT_s]$ is a realization of a  discrete,  stationary, white  Gaussian process, with $E\{\mathbf{w}[k]\}=0,\;
E\{\mathbf{w}[k]\mathbf{w}[k]^T\}=\sigma^2_w I
\;\forall{k}$; the samples of 
 discrete sequences  $b[kT_s]$ are realizations of a zero mean Gaussian random distribution with variance $\sigma^2_b$; and  $s^{(h)}[kT_s], h=0,\cdots H-1$ are drawn from a zero mean unit variance Gaussian random variable.

Once SoGs samples $\mathbf{y}[kT_s]$ are obtained, we estimate the adjacency matrix. There are many state-of-the-art methods to perform the estimation, such as spectral coherence \cite{carter1987coherence}, imaginary coherence \cite{nolte2004identifying}, phase-locking value which differently characterize brain interactions \cite{cattai2018characterization}of the signals at two nodes $i,j$ as:
\begin{equation}
    C_{ij}(\omega_k)=\dfrac{\vert \hat{P}_{ij}(\omega_k)\vert}{\sqrt{\hat{P}_i(\omega_k)\cdot \hat{P}_j(\omega_k)}}
    \label{eq:coh}
\end{equation}
In Eq. \eqref{eq:coh}, $\hat{P}_i(\omega_k)$, $\hat{P}_j(\omega_k)$ and $\hat{P}_{ij}(\omega_k)$ are the  the estimated auto-spectra and cross-spectrum of the signals
${y}_i[kT_s]$,
${y}_j[kT_s]$
at the nodes $i$ and $j$, computed at the frequency bin%
\footnote{All the power spectral estimates are computed with Welch method, with $1$s length Hanning windows and overlap of $50\%$.} $\omega_k=\dfrac{2\pi}{N_s} k$. 
 Given $C_{ij}(\omega_k)$ as in Eq. \eqref{eq:coh},  the  adjacency matrix $\hat{\boldsymbol{A}}$,estimated is  averaging across the $N_s$ frequency bins as follows: 
\begin{equation}
    \hat{{A}}_{ij}=\sum_{k=0}^{N_s-1} C_{ij}(\omega_k)
    \label{eq:cohaveragesynth}
\end{equation}

\begin{figure}[ht]
		\centerline{\includegraphics[scale=.45]{{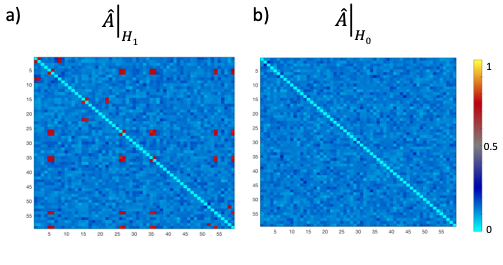}}}
		\caption{Adjacency matrix with synthetic data. $\hat{\boldsymbol{A}}_{ij}[k]$ is represented under $\gHuno$ in a) and under $\gHzero$ in panel b). }
		\label{fig:adj}

\end{figure}

To sum up, our proposed signal model for synthetic data determines a simple graph connectivity under the two hypotheses $\gHuno$ and $\gHzero$.  The model successfully  reproduces a network  characterized by distinguishable connectivity states  in presence of controlled  perturbations. In Fig. \ref{fig:adj}, the estimated adjacency matrix is plotted under the two conditions $\gHuno$ and $\gHzero$ in presence of perturbations. We recognize that under $\gHzero$ there are no links and  $\hat{\boldsymbol{A}}$ fluctuates around zero because of the perturbations. Under $\gHuno$ some connections  exist but their values are affected by the perturbations.

\begin{figure}[ht]
		\centerline{\includegraphics[scale=.45]{{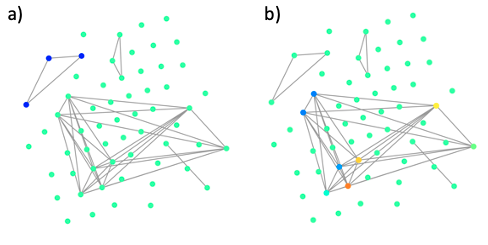}}}
		\caption{Eigenvectors on graph. In panel a) there is the fist eigenvector; in panel b) the $10^{th}$ eigenvector}
		\label{fig:eiggraph}

\end{figure}

Once we have obtained the adjacency matrix estimations under $\gHuno$ and $\gHzero$, we compute the estimated Laplacians as in Eq. \eqref{eq:estlaplacian} and then, we decompose it with its eigenvalues and eigenvectors as in Eq. \eqref{eq:laplestim}. 
In order to recall the eigenvectors' behaviour, we plot in Fig. \ref{fig:eiggraph} the first and the $10^{th}$ eigenvectors on graph under $\gHuno$ hypothesis. We can see that the first eigenvector, in Fig. \ref{fig:eiggraph}(a) perfectly appears smooth on the graph and within  a subset  of linked nodes. Fig. \ref{fig:eiggraph}(b) describes the $10_{th}$ eigenvector  on graph. Here, the eigenvector is mostly smooth, it highlights another community,  but it shows   higher variability than the first eigenvector over linked nodes. 

This model will be used  to test the proposed Laplacian denoising algorithm. To this goal, we randomly produce $20$ repetitions (or trials) for each statistical hypothesis, as i.i.d. realizations of our model with a fixed set of parameters. Different   noise and polarization level will be considered.

\subsection{Sub-space robustness on synthetic data}
 In this sub-section, we compare our laplacian-based filtering based on $\mathcal{U}_{L}\cup \mathcal{U}_{H}$, shortly denoted as $\mathcal{U}_{L\cup H}$, with $\mathcal{U}_{ALL}$, $\mathcal{U}_{L}$ and $\mathcal{U}_{H}$. 
 Specifically, we investigate the robustness of the different sub-spaces with simulated data. 
 
\begin{figure}[ht]
		\centerline{\includegraphics[scale=.38]{{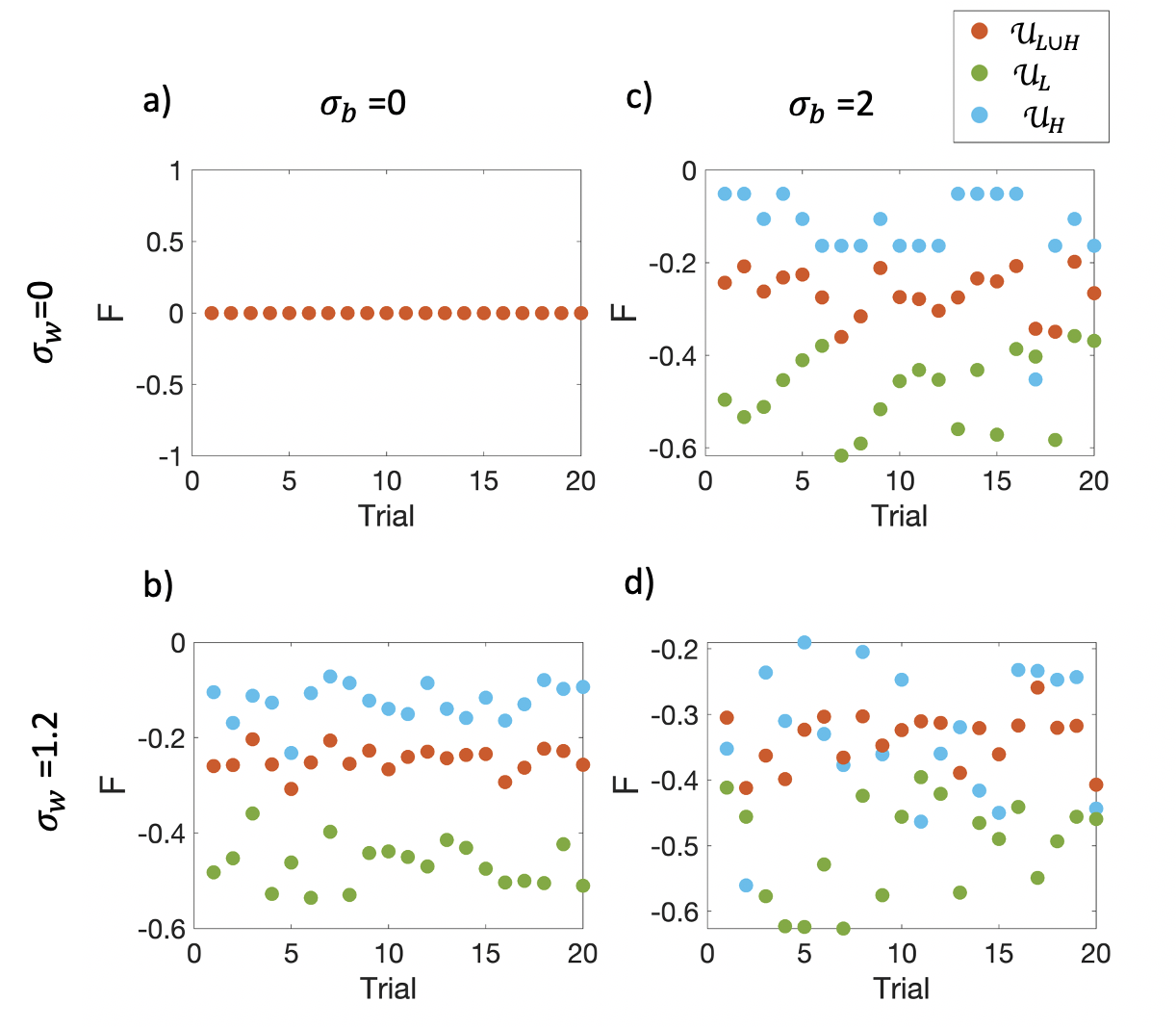}}}
		\caption{Results of Frobenius distance on synthetic data. Several perturbation configuration are represented: in panel a)$\sigma_w$=$0$ and $\sigma_b$=$0$,in panel b) $\sigma_w$=$1.2$ and $\sigma_b$=$0$,in panel c) $\sigma_w$=$0$ and $\sigma_b$=$2$ and in panel d) $\sigma_w$=$1.2$ and $\sigma_b$=$2$. In the different colors (in the legend) we represent the different sub-spaces.}
		\label{fig:frobdist}

\end{figure}

With the final goal of measuring the sub-space robustness, for each sub-space we take into account   two cases, namely  the  absence  and the presence of perturbation, to which we refer to as the ground truth (GT) and the noisy cases, respectively. Each  ground truth sub-space is compared to different noisy cases, corresponding to  $\sigma_w={0, 1.2}$ for noise and $\sigma_b={0, 2}$ for polarization.
To quantify sub-space robustness on synthetic data, we firstly measure the Frobenius subspace distance $F$ \cite{baksalary2014subspace} between the GT case and the noisy configurations, varying the perturbation levels%
\footnote{ For each sub-space configuration, we compute the Frobenius distance $F$ between its noisy and GT versions. See \textbf{Definition 2} in \cite{baksalary2014subspace} for the mathematical formulation.}. 
Results of this analysis are in Fig. \ref{fig:frobdist}. We  plot  $F$ versus the  trial for the subspace $\mathcal{U}_{ H}$ (red), the subspace $\mathcal{U}_{L}$ (green), and the subspace $\mathcal{U}_{L\cup H}$ (blue)  in several perturbation conditions. In Fig. \ref{fig:frobdist}(a) we have the zero perturbation case, in which, not surprisingly, $F=0$ for every filter and every trial. With a gradual increase of perturbations (i.e.  noise only case in Fig.\ref{fig:frobdist}(b) or polarization only case in Fig.\ref{fig:frobdist}(c)), the most favorable case is with $\mathcal{U}_{H}$ for almost every trial. When perturbations dramatically increase Fig.\ref{fig:frobdist}(d), performances decrease in particular for $\mathcal{U}_{H}$ filter. In this figure, we do not report results for $\mathcal{U}_{ALL}$ case because $F=0$ for all the trials with all the eigenvectors.

It is then clear that the eigenvectors in $\mathcal{U}_{H}$ are significantly robust. 
This is not surprising, since in classical signal processing $\mathcal{U}_{H}$  larger eigenvectors are  used because of their advantages in signal-to-noise-ratio (SNR).
Still, the subspace $\mathcal{U}_{L\cup H}$ maintains the robustness, while being  relevant to describe the inherent topology of the graph. In the next results, we show that the Laplacian denoising leveraging the subspace $\mathcal{U}_{L\cup H}$  leads to better distinguishable  connectivity states in absence and in presence of perturbation.   
\begin{figure*}[ht]
		\centerline{\includegraphics[scale=.4]{{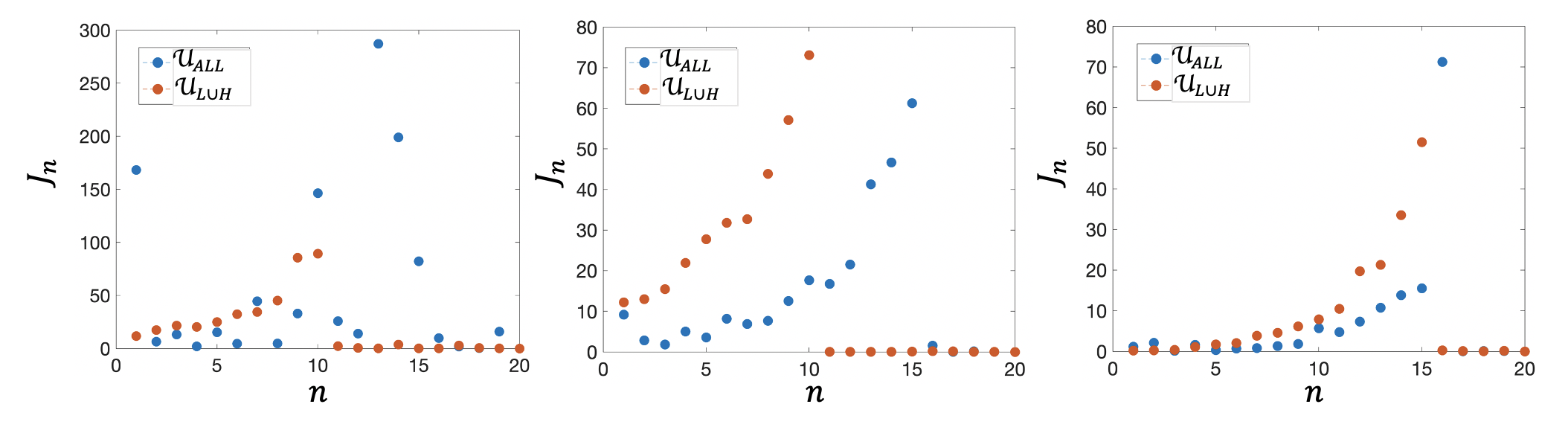}}}
		\caption{Results of J-divergence analysis on synthetic data. Several perturbation configuration are represented: in panel a) $\sigma_w$=$0$ and $\sigma_b$=$0$, in panel b) $\sigma_w$=$1.2$ and $\sigma_b$=$0$, in panel c) $\sigma_w$=$1.2$ and $\sigma_b$=$2$. In the different colors (shown in the legend) we represent the different sub-spaces for the filtering.}
		\label{fig:Jsim}
\end{figure*}
\subsection{J-divergence computation on synthetic data}
Finally, we test the ability to separate graph Laplacians under the hypotheses $\gHuno$ and $\gHzero$. The J-divergence analysis presented in section \ref{sec:Jdiv} ends with a measure of the statistical distance $J$ between two states. Here, the analysis of separability between two states is applied to graph laplacian in simulated scenario with the goal of comparing the discriminant ability of our denoising method with respect to the other sub-space configurations, i.e. $\mathcal{U}_{ALL}$, $\mathcal{U}_{L}$ and $\mathcal{U}_{H}$.

Here the two conditions are  the two hypothesis $\gHuno$ and $\gHzero$ and we perform simulations for several perturbation levels. In every case, we compute the total $J$ as a measure of statistical distance between the two conditions and we also evaluate the marginal $J_n$ as measure of the contribution of each n-variable to the final separability. Table \ref{table:table_Jsim} collects J-divergence values for several perturbation sets and for different sub-spaces. Results show that in absence of perturbations the most favorable case is $\mathcal{U}_{ALL}$. This result is intuitive because in absence of perturbations there is no reason why reduced sub-spaces should better discriminate. Interestingly, increasing perturbations (i.e. noise and polarization), the most favorable case is $\mathcal{U}_{L\cup H}$, which gives the highest $J$, i.e. the best separability. It means that graph laplacian denoising through $\mathcal{U}_{L\cup H}$ preserves the highest separability between the two hypothesis, even in presence of strong perturbation. Fig \ref{fig:Jsim}, shows $J_n$ behavior as function of the first $20$ variables only for $U_{ALL}$ and $U_{L\cup H}$ cases. These representations make clear the contribution of $n$ variables to the final $J$. From  Fig \ref{fig:Jsim} we recognize that increasing perturbations, variables in $U_{L\cup H}$ generally give higher $J_n$ contributions compared to $U_{ALL}$.
\begin{table}[ht]
		\centerline{\includegraphics[scale=.4]{{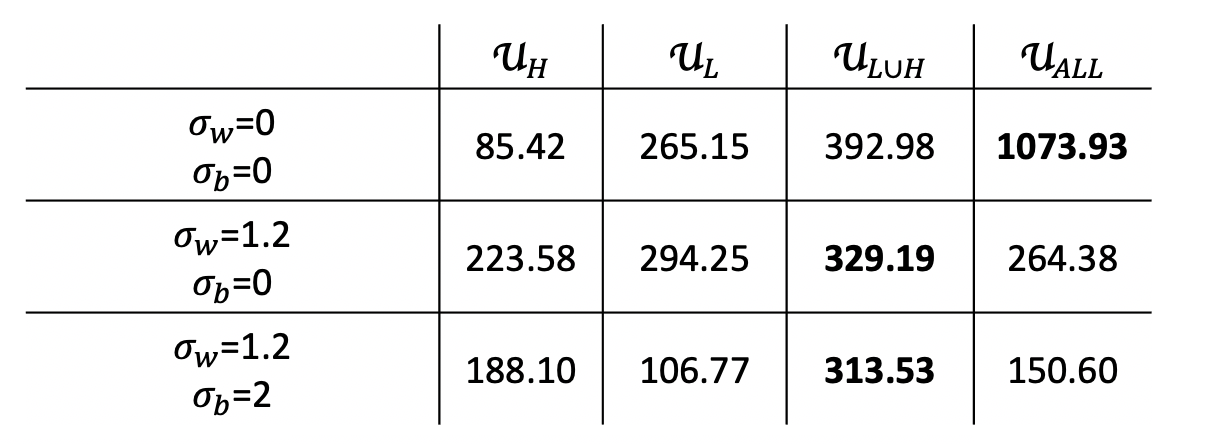}}}
		\caption{J-divergence values on synthetic data. We report in bold characters the highest J-divergence value for each perturbation configuration.}
		\label{table:table_Jsim}
\end{table}

Our results with synthetic data show that in presence of perturbations our laplacian filtering succeeds in distinguishing graphs under two conditions. This conclusion remains true if the system is perturbed by noise but also if there is an artefact of different nature, i.e. a common artifact that we indicated as polarization and which can represent real phenomena.

\section{Real BCI measurements}
 \label{sec:real}
In this section, we present experimental results of our graph laplacian filtering on real data, recorded during motor-imagery BCI experiments. In this case the $\gHuno$ and $\gHzero$ hypothesis directly correspond to he hypothesis that subject performs motor imagery ($\gHuno$) or he/she is in resting state ($\gHzero$).  

\subsection{Experimental Protocol and Preprocessing }

The study was conducted on twenty healthy subjects (aged $27.60 \pm 4.01$ years, $8$ women), all right-handed.  All the subjects, which did not present any disorder, received financial compensation for their participation and they signed a written informed consent. The ethical committee CPP-IDF-VI of Paris approved the experimental protocol.
During the BCI experiments, every subject was in front of a screen with a target. Subjects were instructed to perform right hand - motor imagery task  when the target was up, while remaining at rest when the target was down \cite{wolpaw2003wadsworth}.
A $74$-channel system was used to record EEG data in a standard $10$-$10$ configuration. The reference for EEG were set to mastoid signals; the ground electrode was located on the left scalpula; and impedences were lower than 20 kOhms. Sampling frequency for EEG recordings was 1 kHz, and then downsampled to 250 Hz. 
For each subject, recorded sequences have been segmented to obtain $N_T$ trials of motor imagery and $N_T$ trials of resting state. The total length of each trial was $5$s.

EEG data analysis was preceded by a pre-processing stage. Specifically, an Independent Component Analysis (ICA) was performed to eliminate artifacts, such as ocular and cardiac signals \cite{delorme2007enhanced}; in particular, the Infomax Algorithm \cite{bell1995information} was implemented with Fieldtrip toolbox \cite{oostenveld2011fieldtrip}. 

\subsection{J-divergence of brain connectivity states}

Here, we perform the J-divergence analysis on real motor-imagery data. To this aim, we take EEG signals from one subject and  $N_T$ trials for $\gHuno$ and $N_T$ trials for $\gHzero$, with $N_T=20$. We use spectral coherence to build the connectivity matrix, as in Eq. \eqref{eq:coh}. Then, the estimated adjacency matrix $\hat{\boldsymbol{A}}$ is computed as in Eq. \eqref{eq:cohaveragesynth}, and thanks to the Eq. \eqref{eq:estlaplacian}, we can derive $\hat{\boldsymbol{L}}$. As for synthetic data, we compute the filtered graph laplacian $\tilde{\boldsymbol{L}}$ with the subset $\mathcal{U}_{L\cup H}$. We compare results on real data with $\mathcal{U}_{ALL}$, $\mathcal{U}_{H}$, $\mathcal{U}_{L}$. In each case, we compute J-divergence $J$ as in Eq. \eqref{eq:jdiv} and the marginal contribution $J_n$  associated to the $n$-th variable as in Eq.\eqref{eq:JDivCosBis}.

\begin{table}[ht]
    \centering
    \centerline{\includegraphics[scale=.5]{{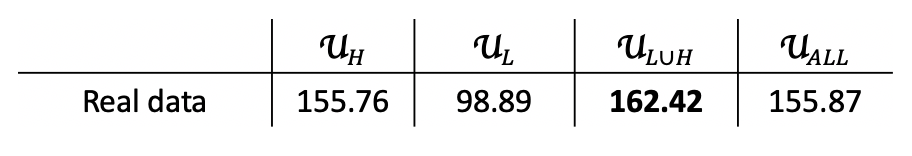}}}
    \caption{J-divergence values on real data. We report in bold characters the highest J-divergence value.}
    \label{table:Jdiv_real}
\end{table}

\begin{figure}[ht]
		\centerline{\includegraphics[scale=.37]{{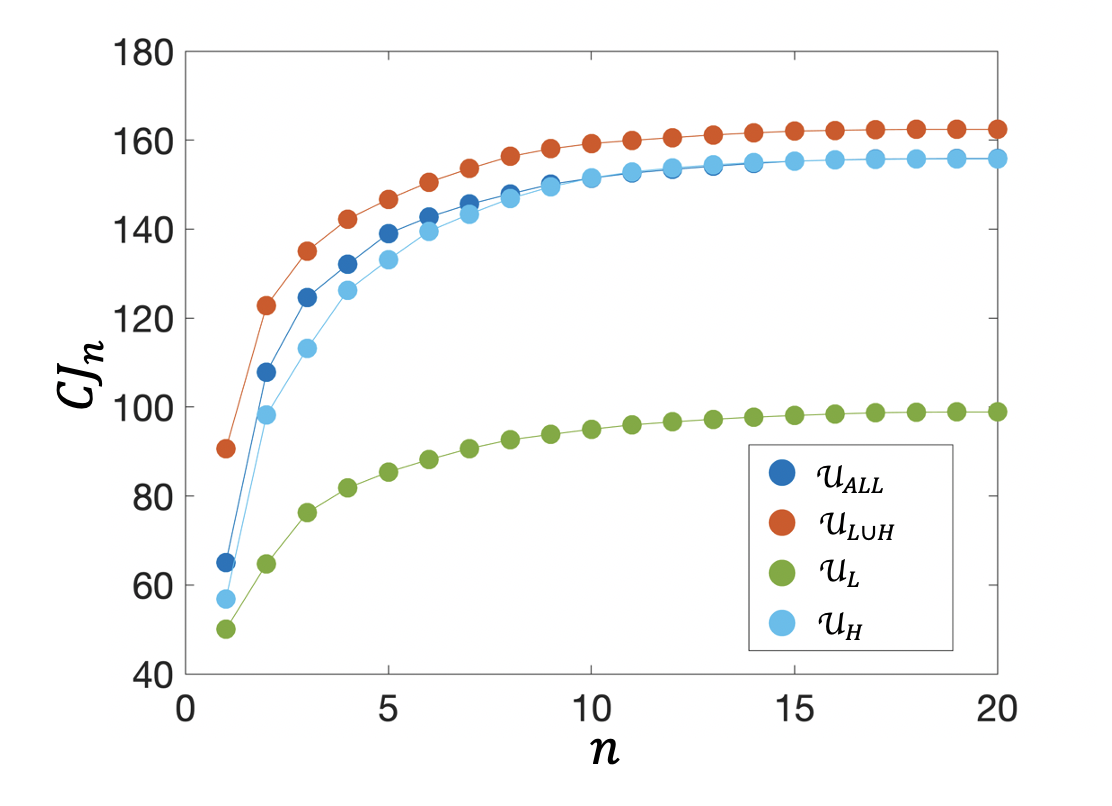}}}
		\caption{Results of J-divergence analysis for real data. We report the $CJ_n$  in Eq.\eqref{eq:CJ} as function of the involved variables. In the different colors (shown in the legend) we represent the different sub-spaces used for filtering.}
		\label{fig:cum_real}
\end{figure}
In Table \ref{table:Jdiv_real}, we report J-divergence results for each sub-space configuration. Comparing all the methods, the highest $J$ value relates to $\mathcal{U}_{L\cup H}$ case. This result is very  important because it means that the sub-space $\mathcal{U}_{L\cup H}$ is suitable to separate real EEG data and it is useful to correctly detect the subject mental state. 

With the aim of understanding the contributions of different variables, we firstly  compute the $J_n$ marginal contributions to obtain a weight to each variable in the transformed domain. Then, we compute the cumulative sum of the first $n$ variables. Once the $J_n$ vector is sorted, we evaluate the cumulative J-divergence $CJ_n$ to investigate the impact of the variables to the total J-divergence, as follows:
\begin{equation}
    {CJ}_n = \sum_{k=1}^{n}J_k
\label{eq:CJ}\end{equation}
Results in Fig. \ref{fig:cum_real} show that the cumulative sum of the first $20$ variables, is always higher for $\mathcal{U}_{L\cup H}$ than all the other sub-space configurations. In other words, if a given number of variables are retained, the overall achieved J-divergence is always larger using the proposed denoising algorithm. This  confirms the contribution  of the proposed Laplacian denoising to  discrimination of the two mental states.

\subsection{{Scoring of Laplacian coefficients in $\beta$ band}}

Given  that the filtering with $\mathcal{U}_{L\cup H}$ enables a better discrimination between motor imagery and resting state, 
we  now exploit the above introduced scoring procedure to determine, based on  an information theoretic grounded  criterion, which connectivity coefficients mostly contribute to separate the resting and motor imagery states. 

To proceed, it is important to underline that the brain response to motor tasks in general is not uniform across the frequencies, but it is mostly evident in $\alpha$ ($8$-$13$ Hz) or $\beta$ ($14$-$29$ Hz) band \cite{meirovitch2015alpha}, depending on the subject.

As a proof of concept, we show results in $\beta$ band, but in a training BCI scenario, the frequency band of interest can be tuned according to the subject response. Here, we filter the connectivity matrix in the selected frequency band, as follows: 
\begin{equation}
    \hat{{A}}_{ij}=\sum_{\omega_k/T_s\in \beta } C_{ij}(\omega_k)
    \label{eq:cohaveragereal}
\end{equation}

\begin{figure*}[ht]
		\centerline{\includegraphics[scale=.43]{{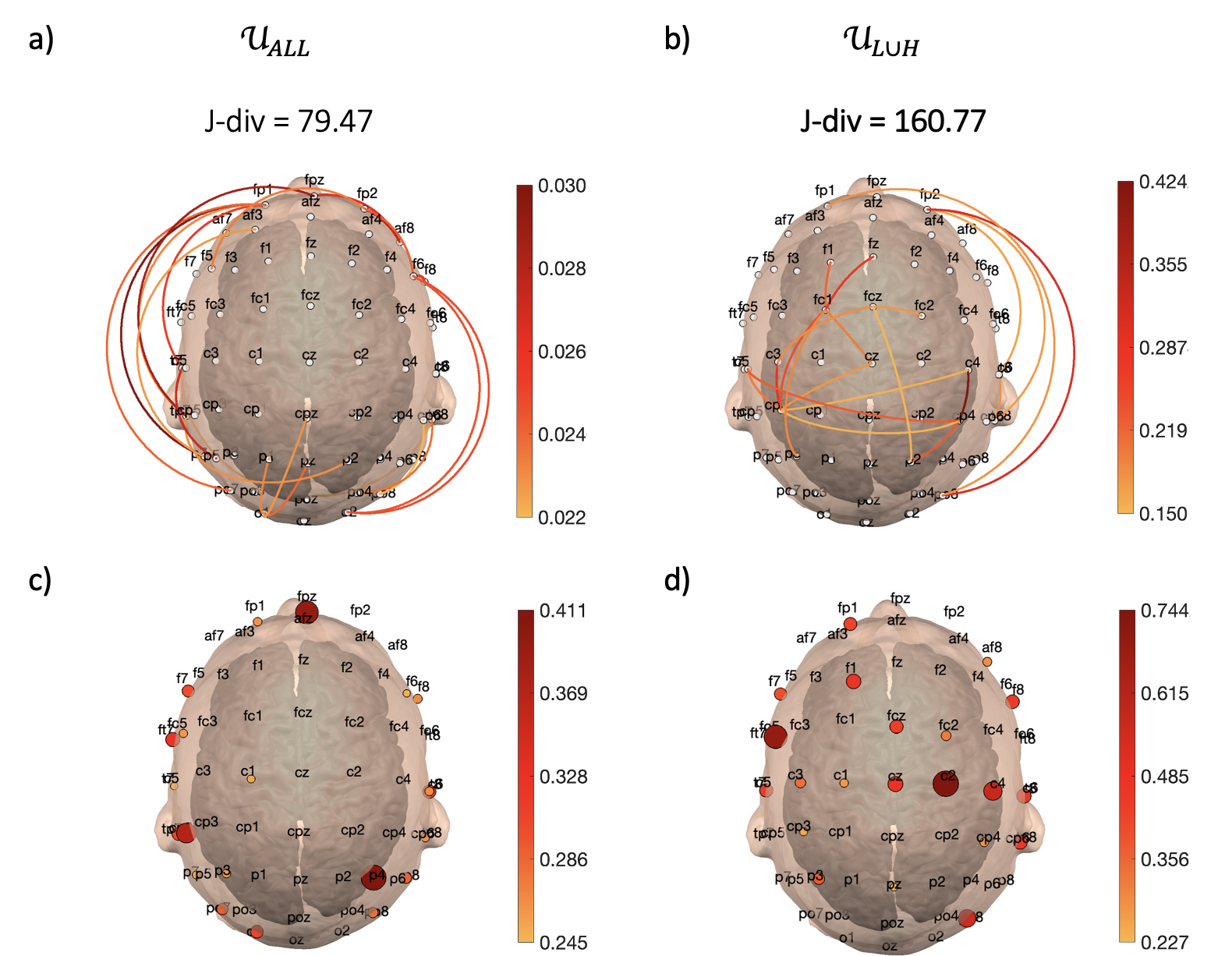}}}
		\caption{Results of score computation for real data. We report results for $\mathcal{U}_{ALL}$ subspace filtering in panels a,c);for $\mathcal{U}_{L\cup H}$ subspace filtering in panels b,d). In the first line, score values refer to links (i.e. extra-diagonal elements) and in the second line, they relate to elements in the principal diagonal (i.e. nodes). { For sake of clarity, in all the  figures we plot the $20$ nodes or links with highest score.}}
		\label{fig:score}
\end{figure*}

Thereby, having stated that the denoising based on  $\mathcal{U}_{L\cup H}$ sub-space provides best results in separating Laplacians under $\gHuno$ and $\gHzero$, we restrict the analysis  to $\mathcal{U}_{L\cup H}$ and $\mathcal{U}_{ALL}$ for score analysis in $\beta$ band.
We compute the scores as described in section \ref{sec:Jdiv}(A) and we report the score results in Fig. \ref{fig:score}.  
In the first row, we collect the results referring to extra diagonal elements of $\tilde{L}$, i.e. links, and in the second row, we report results for diagonal elements, i.e. nodes weights. Besides, on the left and right columns we provide the results achieved without and with application of the proposed Laplacian denoising.

The first interesting observation is that, also  when the analysis is restricted to the $\beta$ band, application of the proposed denoising improves the J-divergence of the observed connectivity states  (from $79.47$ to $160.77$).

As far as the score analysis is concerned, different remarks are in order.

{In Fig. \ref{fig:score}(c-d), the score associated to nodes weigths (Laplacian diagonal elements ) is represented. The score range in absence of denoising is smaller, ie. maximum values are $0.4$ and $0.7$. Furthermore, the scoring  obtained without denoising  is larger on nodes located in frontal, temporal or parietal area, such as $FP_Z$ and $P_4$. After denoising, the scores are more pronounced on sensory-motor areas, and we are able to pinpoint some more relevant nodes, such as $C_2$ and $FC_5$.}
{Let us now analyze   the score associated to links' weigths (Laplacian extra- diagonal elements). In absence of denoising, recognizing contributions of different brain areas is difficult because all the link weights are generally low, ie. between $0$ and $0.035$. 
Besides, we can observe that  the $20$ links with highest score do not involve sensory-motor nodes. On the contrary, when the denoised Laplacian is considered, link scores achieve  higher values, i.e. $0.42$,  the strongest links are localized in sensory-motor areas, and links connecting contro-lateral motor areas, such as $CP_3-C_3$ rank highest.}

Thereby, the scoring based on the denoised Laplacian provides a mean for  analysis and interpretation of the observed connectivity states.

\subsection{Fast estimation of Laplacian coefficients in \textbeta{}  band}

BCIs  aim to provide real time interaction between the subject and the interface \cite{wolpaw2003wadsworth,wolpaw2012brain}; thereby, reducing the observation time $T_{oss}$ for Laplacian estimation is beneficial for potential applicability to online motor-imagery BCI.  
With this application framework in mind, we  test the  Laplacian denoising  when the observation time window length is reduced to $T_{oss}=1s$.  In the following we  consider a moving window of length $T_{oss}=1s$ and shift  it by $m\Delta t, m=0\cdots M-1$, with $M=9$ $\Delta t=0.5$s, so as to analyze the total available recording length of $5$s over  nine $50$\% overlapping temporal intervals. \cite{shenoy2006towards}.

For each of the $20$ subjects of the experimental study,   we compute the spectral coherence on the $m$-th temporal interval, $m=0\cdots M-1$  as in Eq. \eqref{eq:coh}  both for resting  ($\gHuno$) and motor imagery ($\gHzero$) state.  Then we derive the conditional ($\gHuno, \;\gHzero$) estimated adjacency matrix $\hat{{A}}$ as in Eq.\eqref{eq:cohaveragereal},  the estimated graph laplacian $\hat{{L}}$ as in Eq. \eqref{eq:laplestim}, and its  denoised version   $\tilde{{L}}$  as in Eq.\eqref{eq:lapleigdenoise}. Then, we evaluate the J-divergence between the two hypotheses as in Eq. \eqref{eq:JDivCosBis}. Finally, we average the  $J$ obtained on the $m$-th window $m=0\cdots M-1$ in each time-interval across subjects. For comparison sake, we  repeat the above computations in absence of denoising. 
 
\begin{figure}[ht]
		\centerline{\includegraphics[scale=.45]{{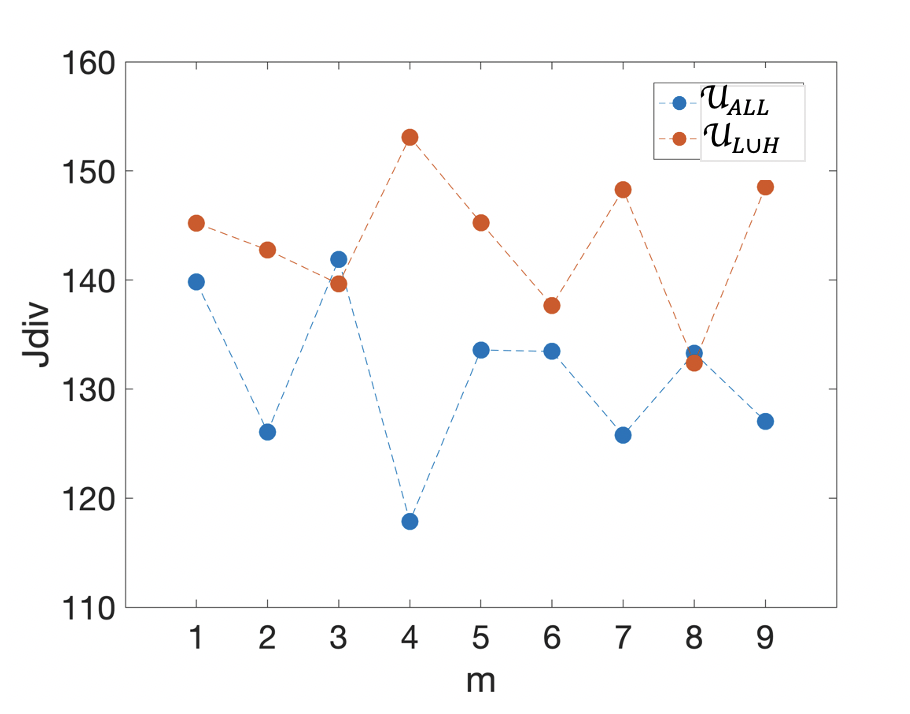}}}
		\caption{Results of  J-divergence analysis over a moving window on real data. We plot the  J-divergence  over $M=9$, $1$s long, time intervals  with  $50$\% overlapping, versus the time interval index. The   J-divergence is computed  in $\beta$ band and averaged across subjects. }
		\label{fig:TVJdiv}

\end{figure}

 Fig. \ref{fig:TVJdiv} reports results of this analysis by plotting  the J-divergence, averaged across subject, as a function of the time window index $m,\;m=0\cdots M-1$.  Our findings show that in the majority of the  considered time intervals, i.e. on $7$ intervals out of $9$,   the denoised Laplacian  $\tilde{{L}}$ , leveraging the $\mathcal{U}_{L \cup H}$ subspace, leads to higher J-divergence than the  estimated Laplacian $\hat{{L}}$ ($\mathcal{U}_{ALL}$ subspace). This result is really interesting because it shows that, even with short time-interval, our method succeeds in separating the two mental states. 
The above findings on real EEG data show that the proposed    Laplacian denoising applies also on   short time-windows and  improves the potential to correctly detect motor imagery state. This paves the way to application of the proposed Laplacian denoising to real BCI applications.

\section{Conclusion and further work}
 \label{sec:concl}
This work has  proposed  a Laplacian denoising algorithm for the purpose of graph connectivity states detection. A novel formulation of the Jensen divergence has been derived. The J-divergence formulation is used to quantify the performance of the denoising algorithm,  as well as to attribute a score to the Laplacian coefficients in terms of their contribution to the connectivity states separability. The Laplacian denoising algorithm performances are  assessed by numerical simulations on synthetic data. Furthermore,   the Laplacian denoising  algorithm has been applied to real EEG data acquired within motor imagery BCI experiments.  The proposed Laplacian denoising improves the separation of the two mental states of motor imagery and resting state, even under restrained observation time intervals. Besides, the J-divergence  based scoring  sheds light over the contribution of different  connectivity coefficients to motor imagery state detection. Thereby,  the proposed approach  is promising for the robust detection of  connectivity states while being appealing     for implementation in real-time   BCI applications.

\begin{appendices}
\section{Theorem $1$}
Let us consider the problem of  binary classification of Gaussian variables \(
 \gHzero: \gvx \sim \gNormalLaw{\gvzero}{\gvI}
\), \(
\gHuno: \gvx \sim \gNormalLaw{\gveta}{\gD}
\), corresponding to the uncommon mean, uncommon covariance case,  by means of  the LLRT formulation in Eq.\eqref{eq:LRTsimple}. By simple algebraic manipulation, we recognize that the test $\gLRTD(\gvx)\gvrv{\gHuno}{\gHzero} t'$ corresponds to:
\begin{equation}
\begin{split}
\gLRTD'(\gvx) 
  &
= \underbrace{\sumdneqone\dfrac{\gdnsq-1 }{\gdnsq}  \left|\gxn + \frac{\getan}{\gdnsq-1} \right|^2}_{\textit{\(\gP\) quadratic terms}} 
+ 
 \underbrace{\sumdeqone\!\!2{\getan\cdot \gxn}}_{\textit{\(\gN\gmeno\gP\) linear terms}}
 \gvrv{\gHuno}{\gHzero} t''
\end{split}
\end{equation}
with \(t''\!=\!t'  +  \sumdneqone {|\getansq|}\left[\gdnsq  \left(\gdnsq-1\right)\right]^{-1}
\).

Let us consider the linear-quadratic observation space \(\Xi\) of  the  \(N\)-dimensional  random vector  
\( {\gvxi}\ggdef\left[\gxi_{1}\dots \gxi_\gN\right]\ttt\) defined as (see Eq. \eqref{eq:x2xi})
\begin{equation}
\begin{split}
\gxi_n&=  \left(\gxn + \frac{\getan}{\gdnsq-1} \right)^2
\\
&=\gxn^2+2 \;\gxn\;
\;\frac{\getan}{\gdnsq-1}+
\left(
\frac{\getan}{\gdnsq-1} \right)^2;
\gn=1,\dotsc, \gP\\
\gxi_n&=\gxn, \;\gn=\gP+1,\dotsc, \gN
\end{split}
\end{equation}

In the space \(\Xi\) the LLRT $\gLRTD'(\gvx)\gvrv{\gHuno}{\gHzero} t''$
rewrites as follows:
\begin{equation}
\sum\limits_0^{N-1} \gcLLRTn \xi_n ={\gvcLLRT\hhh\cdot\gvxi} 
\gvrv{\gHuno}{\gHzero} t''
\label{eq:LLRxi}
\end{equation}
where  the elements of  \( \gvcLLRT \ggdef \left[\gcLLRTuno,\dotsc,\gcLLRTN\right]\ttt \)  are:
\begin{align}
\gcLLRTn & \ggdef 
\begin{cases}
\dfrac{\gsigmaysqn-\gsigmaysqnm}{\sigma_n}
& \text{for \(\gn=1,\gP\)}\\
2\getayn & \text{for \(\gn=\gP+1,\gN\)}
\end{cases}
\label{eq:cLRTnBis}
\end{align}

With these positions, 

\begin{equation}
\begin{split}
 J&\ggdef \E{\gLRTD(\gvx)|\gHuno} - \E{\gLRTD(\gvx)|\gHzero}  
 \\&=
 \E{\gLRTD'(\gvx)|\gHuno} - \E{\gLRTD'(\gvx)|\gHzero}
 \\&=
 \sum\limits_{n=0}^{N-1}
 \gcLLRTn 
 \left(\E{\xi_n |\gHuno} - \E{\xi_n |\gHzero}\right)
\end{split}
\end{equation}   
By computing the above expectations it can be straightforwardly shown that the $n$-th term $\gcLLRTn
 \left(\E{\xi_n |\gHuno} - \E{\xi_n |\gHzero}\right)$ of the above sum equals  to  \begin{equation} \begin{split}
    J_n&^{(\sigma_n,\eta_n)}
 =\dfrac{\gsigmaysqn-\gsigmaysqnm}{\sigma_n}
 \left(\sigma_n^2+\eta_n^2 +2\dfrac{\eta_n^2}{\sigma_n^2-1}-1\right)
 \\&=
 \left(\sigma_n-\sigma_n^{-1}\right)
 \left[
 \left(\sigma_n-\sigma_n^{-1}\right)
 +
 \dfrac{\eta_n^2}{\sigma_n}
 \dfrac{\sigma_n^{2}+1}{\sigma_n^{2}-1}
 \right]
  \\&=
 \left(\sigma_n-\sigma_n^{-1}\right)^2
 \left[
 1
 +
 \dfrac{\eta_n^2}{\sigma_n}
 \dfrac{\sigma_n^{2}+1}{
 \left(\sigma_n-\sigma_n^{-1}\right)
 \left(\sigma_n^{2}-1\right)}
 \right]
 \\&=
 \left(\sigma_n-\sigma_n^{-1}\right)^2
 \left[
 1
 +
 \dfrac{\eta_n^2}{\sigma_n}
 \dfrac{\sigma_n+\sigma_n^{-1}}{
 \left(\sigma_n-\sigma_n^{-1}\right)^2}
 \right],
 n=1,\cdots P
 \\
 J_n&^{(\eta_n)}=2{\eta_n^2},\;n=P,\cdots N-1.
\end{split}\end{equation}
and 
\begin{equation}
J_n^{(\eta_n)}=2{\eta_n^2},\;n=P,\cdots N-1.
\end{equation}
QED.
\end{appendices}

\section*{Acknowledgements}
FD acknowledges support from the Agence Nationale de la Recherche through contract number ANR15$-$NEUC$-0006-02$; and the European Research Council (ERC) under the European Union's Horizon $2020$ research and innovation programme (grant agreement No. $864729$). TC acknowledges Juliana Gonzalez-Astudillo for useful discussions and suggestions.

\bibliography{bib_Jdiv}
\bibliographystyle{ieeetr}

\end{document}